\documentclass[traditabstract]{aa}
\usepackage{graphicx}
\usepackage{amssymb}
\usepackage{amsfonts}
\usepackage{txfonts}
\usepackage{natbib}
\bibpunct{(}{)}{;}{a}{}{,}

\begin{document}
\title{Radially extended kinematics and stellar populations of the massive 
ellipticals NGC1600, \object{NGC4125} and NGC7619.}
\subtitle{Constraints on the outer dark halo density profile.}
\titlerunning{The outer halos of \object{NGC1600}, \object{NGC4125} and 
\object{NGC7619}}
\author{S. B. Pu\inst{1,2,3}
 \thanks{e-mail:pushibi@ynao.ac.cn}
  \and   R. P. Saglia\inst{3,4}
  \and   M. H. Fabricius\inst{3,4}
  \and   J. Thomas\inst{3,4}
  \and   R. Bender\inst{3,4}
  \and   Z. Han\inst{1,5}
         }
 \institute{National Astronomical Observatory/Yunnan Observatory, the Chinese Academy of
 Sciences, Kunming 650011, China
 \and Graduate University of Chinese Academy of Sciences, Beijing 100049, China
 \and Max-Planck-Institut f\"{u}r
extraterrestrische Physik, Postfach 1312, 85741 Garching, Germany
  \and Universit\"{a}tssternwarte, M\"{u}nchen,Scheinerstra$\beta$e 1, D-81679 M\"{u}nchen, Germany
  \and Key Laboratory of the Structure and Evolution of Celestial Objects, Chinese Academy of Sciences, Kunming 650011, China
 }
 \abstract {We present high quality long slit spectra along the major
   and minor axes out to 1.5-2 $R_e$ (14-22 kpc) of three bright
   elliptical galaxies (\object{NGC1600}, \object{NGC4125}, \object{NGC7619}) obtained at the
   Hobby-Eberly Telescope (HET). We derive stellar kinematic profiles
   and Lick/IDS indices ($H\beta,Mgb,Fe_{5015},Fe_{5270},Fe_{5335},
   Fe_{5406}$). Moreover, for \object{NGC4125} we derive gas kinematics and
   emission line strengths. We model the absorption line strengths
   using Simple Stellar Populations models that take into account the
   variation of $[\alpha/Fe]$ and derive ages, total metallicity and
   element abundances. Overall, we find that the three galaxies have
   old and $[\alpha/Fe]$ overabundant stellar populations with no
   significant gradients. The metallicity is supersolar at the center
   with a strong negative radial gradient. For \object{NGC4125}, several pieces
   of evidence point to a recent dissipational merger event.  We
   calculate the broad band color profiles with the help of SSP
   models.  All of the colors show sharp peaks at the center of the
   galaxies, mainly caused by the metallicity gradients, and agree
   well with the measured colors.  Using the Schwarzschild's
   axisymmetric orbit superposition technique, we model the stellar
   kinematics to constrain the dark halos of the galaxies. We use the
   tight correlation between the Mgb strength and local escape
   velocity to set limits on the extent of the halos by testing
   different halo sizes. Logarithmic halos -- cut at 60~kpc --
   minimize the overall scatter of the Mgb-$V_{esc}$ relation. Larger
   cutoff radii are found if the dark matter density profile is
   decreasing more steeply at large radii. }

   \maketitle

\section{Introduction}
\label{sec_intro}
One of the most important remaining questions in contemporary
astrophysics concerns the formation and evolution of stellar
populations in elliptical galaxies.\footnote{Tables 3, 4 and 6 are
  only available in electronic form at the CDS via anonymous ftp to
  cdsarc.u-strasbg.fr (130.79.128.5) or via
  http://cdsweb.u-strasbg.fr/cgi-bin/qcat?J/A+A/}
  One can think of two globally
different formation scenarios: the classic
\emph{monolithic-collapse} model for the formation and evolution of
early-type galaxies \citep{1972ApJ...178..319T,1975MNRAS.173..671L,
1996A&A...311..361T} and the \emph{hierarchical merging} scenario
\citep{1978MNRAS.183..341W,1993MNRAS.264..201K}. In the former,
early-type galaxies build most of their stars during a single event
in the early universe while in the latter a substantial fraction of the stellar
populations is formed from multiple mergers and the accretion of smaller
objects over an extended period. 

Statistically, the extremely small scatter of
the observed color magnitude relation of elliptical galaxies or bright
cluster members suggests that the stellar populations are surprisingly
homogeneous \citep{1992MNRAS.254..601B,1997A&A...320...41K}.  This
uniformity of stellar populations in ellipticals is also supported by
the internal tightness of the so called Fundamental Plane
\citep{1987ApJ...313...42D,1987ApJ...313...59D, 1992ApJ...399..462B,
  1993A&A...279...75S}. These findings are consistent with a \emph{monolithic-collapse}. 
Moreover, the tightness of the \begin{math}Mg-\sigma\end{math} relation 
for massive elliptical galaxies observed in the local universe 
\citep{1993ApJ...411..153B,2007MNRAS.377..759S} but holding up to the 
intermediate redshift 
\begin{math}z\thickapprox 1\end{math} \citep{1997MNRAS.291..527Z,1998ApJ...493..529B}
requires a picture of a short, and highly efficient star formation 
process at high redshift and passive evolution since then.

In contrast, a large number of elliptical
galaxies are observed to have kinematically decoupled cores and were
explained as the result of dissipational major merging events by star
dominated systems
\citep{1992ApJ...393..484B,1989A&A...217...35B,1998A&A...332...33M}.
Furthermore, boxy galaxies are mostly more massive, more radio-loud,
more strong X-ray emitters, more frequently disturbed, and less
flattened than disky early-type galaxies, suggesting that these two
subclasses reflect different formation mechanisms and evolution history
\citep{1992A&A...258..250B,1989A&A...215..266N}. Recently,
\citet{2000MNRAS.315..184K,2005ApJ...621..673T, 2006MNRAS.370.1213C,
2006AJ....131.1288B,2006MNRAS.370..702C,2010MNRAS.tmp..387R}
showed that early-type galaxies in low density and in high density environments
might exhibit different formation ages, and
\citet{2009A&A...499...47S,2009ApJ...691.1862M} found evidence that
lower mass galaxies have more extended star formation histories.

In standard closed-box models of chemical enrichment, the metallicity is
a function of the yield and of how much gas has been locked after star
formation has ceased \citep{1980FCPh....5..287T}.  Therefore, the
metallicity strongly depends on the dynamical parameters -- the
deeper a potential is, the easier it is to retain gas in the
galaxy. Then, galaxies formed in a \emph{monolithic-collapse}
should have steep radial metallicity gradients
\citep{1976MNRAS.176...31L,1999MNRAS.306..655T}, correlated with the
local escape velocity -- since the latter is a measure of the local depth of the
potential well. On the other hand, major mergers,
which come along with \emph{hierarchical merging} will dilute 
stellar population gradients and their connection with the escape
velocity as a result of violent relaxation
\citep{1980MNRAS.191P...1W}. In \emph{hierarchical merging}, however, particularly 
for high-mass galaxies a significant amount of mass accretion can be due to
minor mergers. If this deposites stellar populations that
originally formed in small-mass halos preferentially in the outer parts of the
final massive ellipticals, then this might enhance stellar-population gradients
and one would again expect deviations from correlations with the local escape
velocity, but in the opposite direction as induced by major mergers.
Accordingly, the observation of stellar population gradients and their connection
to dynamical parameters can give crucial insight into the formation paths of
individual galaxies. 

The comparison of measured line strengths to simple stellar population
models \citep[][hereafter
TMB03]{1994ApJS...95..107W,1999ApJ...513..224V,2003MNRAS.339..897T}
yields an estimate of the age, metallicity and element abundances and
the galaxy formation scenarios, formation time scales and star
formation histories can be investigated in detail from the study of
the kinematic properties and line strength features
\citep{1992A&A...258..250B,1992ApJ...399..462B,2004MNRAS.354..753M}.
\citet{2007MNRAS.374..769S} show that when two bursts are considered, the
metallicity and element abundances derived with such a procedure are
roughly luminosity weighted, while the stellar ages can be biased
towards the ones of the younger component. 

There is a considerable number of works which focus on the study of
kinematic profiles, line strength indices and stellar population
parameters in early-type galaxies
\citep{1992A&A...258..250B,1994MNRAS.269..785B,1995ApJ...448..119F,
  1998A&AS..130..267L,1998A&A...333..419T,2000AJ....119.1645T,2000AJ....120..165T,
  2000A&A...353..917L,2000MNRAS.315..184K,2002MNRAS.330..547T,2005MNRAS.358..813D,
  2007A&A...464..853L, 2008MNRAS.386..715T}.  However, the previous
measurements are concentrated within $R_{e}$/2 of galaxies.  From a
comparison of the stellar parameters within $R_{e}$/8 with these
within $R_{e}$/2 of some early-type galaxies sample,
\citet{1993MNRAS.262..650D,2000AJ....119.1645T,2005MNRAS.358..813D}
found that the outer regions of the elliptical galaxies are slightly
more metal poor than the centers, and that ages are likely to increase
slightly outwards.  The same trends were detected by
\citet{1995ApJ...448..119F}. But still only 1/3 of the star mass is
contained within $R_e$/2.  So far the exact behaviour of the stellar
kinematics and the abundance ratios at larger radii is still poorly
known.

Extended stellar kinematic profiles are also welcome to probe the
dynamics of the outer regions of elliptical galaxies, where dark
matter halos start to dominate the gravitational potential.  Together
with other kinematic tracers, like the radial velocities of planetary
nebulae \citep{2009MNRAS.393..329N} and globular clusters 
\citep{2006MNRAS.373..157B}, 
the density and temperature profiles of
X-ray halos  \citep{arXiv:0911.0678} 
or gravitational lensing effects \citep{2009ApJ...703L..51K},
extended stellar kinematics profiles have been used to measure the
dark matter halos densities and constrain their assembly epoch
\citep{Gerhard01, Tho07, Tho09}.

A question of particular interest is the DM halo sizes. It has been
probed statistically using gravitational lensing, and it can be
addressed, if accurate X-ray observations sample the very outer regions
of ellipticals.  Through gravitational lensing a statistical value for
the cutoff radius of $64^{+15}_{-14}$ kpc for the average cluster
galaxy with a velocity dispersion of 220 km/s has been estimated
\citep{2007ApJ...656..739H}. X-ray measurements, on the other hand, do
not seem to detect the ends of halos \citep{arXiv:0911.0678}.

Here we attempt to improve on the leverage of stellar dynamical
models, that can not constrain the mass profiles accurately beyond the
last kinematic data point
\citep{2005MNRAS.360.1355T,2007MNRAS.381.1672T}, by adding the further
constraint coming from the measured absorption line strengths.   \citet{1990ApJ...359L..41F} 
found that the local color is a function
of the local escape velocity. 
 \citet{1993MNRAS.262..650D} discovered a tight correction
between $Mg_2$ and the local escape velocity $V_{esc}$ in 8 galaxies.
More recently, \citet{2009arXiv0906.0018W} also
reported a tight relation between Mg\emph{b} and $V_{esc}$ in NGC3379
while \cite{2009MNRAS.398.1835S} explored correlations between
stellar population parameters and $V_{esc}$ in the SAURON sample.
   
Dynamical measurements can only constrain the density profile of the
dark matter halo out to radii for which data are available.  In
contrast, as the Mgb line strength is a function of the escape
velocity at a given radius, it probes the gravitational potential
directly rather than its gradient, as it is the case with dynamical
measurements. Therefore, since the gravitational potential depends on
the whole density distribution (in principle out to infinity), the
Mgb-$V_{esc}$ provides a novel tool to constrain the density
distribution of dark matter in the outer regions of galaxies.

For this work we obtained deep long slit spectra of three giant local
galaxies, \object{NGC1600}, \object{NGC4125} and \object{NGC7619} (see Table \ref{tab_prop}).
They were chosen as representative of the class massive ellipticals,
that could be easily fit into the service mode schedule 
\citep{2007PASP..119..556S} of the
Hobby-Eberly Telescope (HET) \citep{1998AAS...193.2602B} at McDonald
Observatory.  These galaxies have been well studied in several works
\citep{1994MNRAS.269..785B,1995ApJ...448..119F}.  The galaxy \object{NGC1600},
has boxy isophotes \citep{2002A&A...391..531R} and does not show
rotation both along major and minor axes. The galaxy \object{NGC4125}, has dust
and ionized gas aligned along the stellar major axis
\citep{1989ApJ...346..653K}; \citet{1994A&AS..105..341G} also found
that the dust lane is inclined by 10\degr with respect to the apparent
major axis; in addition, this galaxy shows strong soft (0.1-2.4~keV)
X-ray and far-infrared ($\lambda$ = 60 $\mu$m) emission
\citep{1998AJ....116.2682C}.  For \object{NGC7619} \citet{2002A&A...391..531R}
found evidence of tidal interactions with the galaxy NGC7626.
Recently, \citet{2006ApJ...636..698F} analyzed \emph{Chandra} data and
obtained that \object{NGC1600} and \object{NGC7619} show a positive temperature gradient
towards the outer parts. On the other hand \object{NGC4125} has a declining
temperature profile.  Our spectra extend out to 1.5 $R_e$ for \object{NGC1600}
and more than 2$R_e$ for \object{NGC4125} and \object{NGC7619}.  This dataset allows us
to constrain the general characteristics of the stellar kinematics and
populations extending to large radii. By combining stellar dynamical
modeling with the measured line strength indices we derive constraints
on the extent of the dark matter halos in these galaxies.

The paper is organized as follows. In Sect. \ref{sec_obsdat} we
describe the observations (Sect.  \ref{sec_obs}) and the data
reduction (Sect. \ref{sec_dat}). In Sect.  \ref{sec_kinlin} we present
the kinematics (Sect. \ref{sec_kin}) and the line strength measurements
(Sect. \ref{sec_lin}). Comments on individual galaxies are given in
Sect. \ref{sec_comments}.  We analyze the Lick indices and derive ages,
metallicities, and $\alpha/Fe$ ratios in Sect.
\ref{sec_stelpop}. Section \ref{sec_SPPmod} describes the models and
the method used; Sect. \ref{sec_SPPresults} presents the related
results; Sect. \ref{sec_colors} derives colors and mass-to-light
ratios.  In Sect. \ref{sec_dynamics} we produce dynamical models of
the galaxies. We discuss the adopted method in Sect. \ref{sec_dynmet},
the resulting models (Sect. \ref{sec_dynres}), the constraints coming
from the correlation between Mg\emph{b} and local escape velocity
(Sect. \ref{sec_vesc}) and the resulting degeneracy between dark halo
cut-off radii and outer density slope (Sect. \ref{sec_deg}). A discussion
of the results and a summary of this work are 
presented in Sect.  \ref{sec_summary}

\begin{table*}
\caption{List of the morphological classifications, 
B band luminosity, luminosity distance, receding velocity and effective
radius of \object{NGC1600}, \object{NGC4125} and \object{NGC7619}. They are taken from 
\citet{1989ApJS...69..763F,2001MNRAS.328..461O} 
and NED(NASA/IPAC Extragalactic Database). \label{tab_prop}}
\begin{center}
\begin{tabular}{lccccccr}
\hline
Name   &Ra  & Dec & Type & log($L_B$/$L_\odot$) & Dis & Vel(km/s)  &$R_e$( arcsec)\\
\hline
\object{NGC1600} &  04h31m39s  &-05d11m10s &E3 &11.03 &64Mpc  &4681 &45   \\
\object{NGC4125}  & 12h08m06s  & 65d10m26s &E6 &10.80 &20Mpc  &1356 &58  \\
\object{NGC7619}  & 23h20m14s  &8d12m22s   &E2 &10.58 &46Mpc  &3760 &32  \\
\hline
\end{tabular}
\end{center}
\end{table*}

\section{Observations and data reduction}
\label{sec_obsdat}

\subsection{Observations}
\label{sec_obs}

Long-slit spectra of \object{NGC1600}, \object{NGC4125} and \object{NGC7619} were collected along
the major and minor axis in the period 2006-2008 using the HET in
service mode and the Low-Resolution Spectrograph (LRS) with the E2
grism \citep{1998AAS...193.1003H}. For \object{NGC1600} and \object{NGC7619} the slit
was centered on the center of the galaxies. For \object{NGC4125}, the center of
the galaxy was moved towards one end of the slit, to be able to probe its outer
region despite of its larger half-luminosity radius.  The wavelength
range covered was 4790 to 5850 {\AA} with a pixel size of
0.73{\AA}. The slit width was 3~arcsec, giving an instrumental
broadening of $\sigma_{inst}$ = 120 $\mathrm{km/s}$. Several 900 sec
exposures were taken at each galaxy position. Moreover, 900~sec
exposures of blank sky regions were taken at regular intervals. The
seeing ranged from 1.2 to 2.6 arcsec.  The resulting summed spectra
probe regions out to nearly 1.5 $R_e$ for \object{NGC1600} and more than 2$R_e$
for \object{NGC4125} and \object{NGC7619}.  The list of the spectroscopic observations
is given in Table \ref{tab_log}.

 \begin{table}
 \caption{Log of spectroscopic observation, MJ = major axis; 
 MN = minor axis; SKY= sky spectrum.\label{tab_log}}
 \begin{tabular}{lllc}
 \hline
 Date       & Objects   & Position  &Seeing(FWHM)\\
 \hline

 2006 Sep 19 & NGC 7619   &MN1,2,SKY1       &1.26\\
 2006 Sep 20 &            &MJ1              &1.97\\
 2006 Sep 27 &            &MN3,4,SKY2       &1.66\\
             &            &MJ2,3            &1.34\\
 2006 Sep 28 &            &MJ4,5            &1.18\\
             &            &MN4,5            &1.42\\
 2006 Oct 29 &            &MN6,7            &1.99\\
 2006 Nov 17 &            &MJ6,7            &2.17\\
 2006 Nov 22 &            &MN8,9            &1.47\\
 2007 Jan 28 &\object{NGC4125}     &MJ1,2,SKY1       &1.64\\
             &            &MN1,2            &1.36\\
 2007 Feb 20 &            &MJ3,4,SKY2       &2.31\\
 2007 Feb 21 &            &MN3,4,5,6 SKY3   &2.54\\
 2007 Feb 22 &            &MJ5,6            &1.62\\
 2007 Mar 14 &            &MJ7,8            &2.26\\
 2007 May 05 &            &MJ9,10,SKY4      &2.21\\
 2007 May 10 &            &MN7,8,SKY5       &1.63\\
 2007 May 11 &            &MJ11,12,SKY6     &1.70\\
 2007 Oct 09 &\object{NGC1600}     &MJ1,2,SKY1       &1.76\\
 2007 Oct 16 &            &MN1,2,SKY2       &1.74\\
 2007 Oct 20 &            &MJ3,4,SKY3       &1.42\\
 2007 Oct 21 &            &MN3,4,SKY4       &2.15\\
 2007 Nov 07 &            &MJ5,6,SKY5       &1.66\\
 2007 Nov 19 &            &MN5,6,SKY6       &1.61\\
 2007 Dec 15 &            &MN7,8,SKY7       &1.92\\
 2007 Dec 16 &            &MN9,10,SKY8      &2.20\\
 2008 Jan 10 &            &MJ7,8,SKY9       &2.29\\
 2008 Feb 04 &            &MN11,12,SKY10    &2.57\\
 2008 Feb 09 &            &MJ9,10,SKY11     &1.49\\
\hline
\end{tabular}
\begin{list}{}{}
\item[$^{\mathrm{a}}$] The exposure time for each slit is 900
seconds.
 \item[$^{\mathrm{b}}$] Position angles of our galaxies.
 \object{NGC1600}: [MJ, MN] are [9, 99]; \object{NGC4125}: [MJ, MN] are [82, 172];
 \object{NGC7619}: [MJ, MN] are [36, 126].
\end{list}
\end{table}

In addition, calibration frames (biases, dome flats and the Ne and Cd
lamps) were taken.  Finally, a number of standard template stars (K-
and G-giants) from the list of \citet{1994ApJS...94..687W} were
observed to calibrate our absorption line strengths against the Lick
system.

\subsection{Data reduction}
\label{sec_dat}

The data reduction used the MIDAS package provided by ESO.  The
pre-process of data reduction was done following
\citet{1994MNRAS.269..785B}.  The raw spectra were bias subtracted,
and divided by the flatfields. The cosmic rays were removed with a
$\kappa-\sigma$ clipping procedure. The wavelength calibration was
performed using 9 to 11 strong Ne and Cd emission lines and a third
order polynomial. The achieved accuracy of the wavelength calibration
is always better than 0.6 {\AA} (rms).  The science spectra were
rebinned to a logarithmic wavelength scale. 

As discussed in \citet[ Fig. 2]{2009arXiv0910.5590S}, we also need to
correct the anamorphic distortion of the LRS.  Using stars that were
trailed several times across the slit, we mapped their traces
orthogonal to the wavelength direction and corrected all science
frames by means of sub-pixel shifting using the mapped distortion.  We
estimate that after the correction the maximal residual distortion is
of the order of 1.2 arcsec.

\begin{figure}
  \begin{center}
    \includegraphics[width=0.48\textwidth]{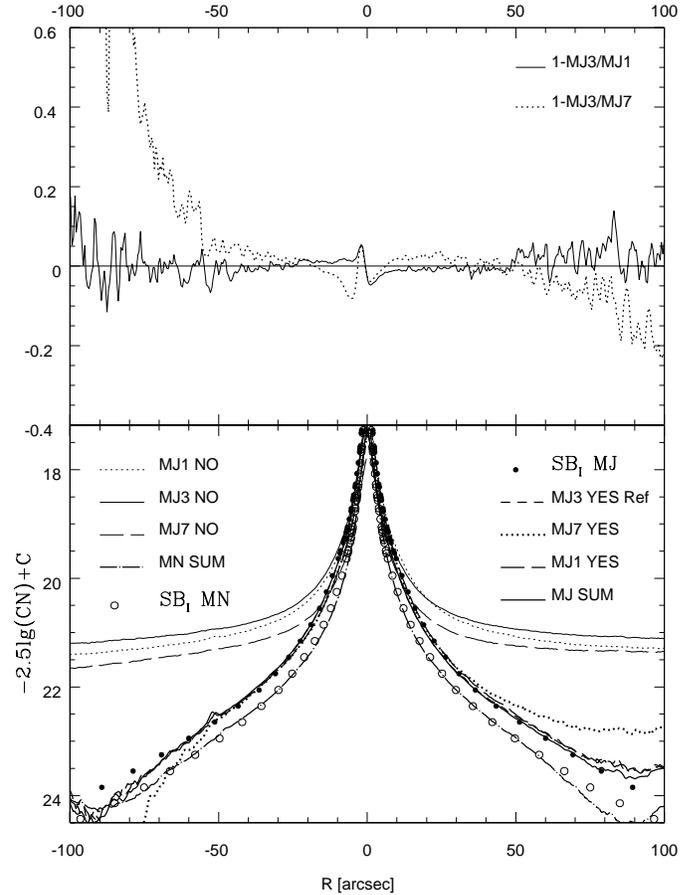}\
  \end{center}
  \caption{Examples of the sky calibration for the major axis spectra
    of \object{NGC7619}. In the bottom of the figure, the X coordinate is the
    distance from the galaxy center in arcsec and the Y coordinate is
    $ -2.5*lg(CN) + 25.7$ (where CN is counts number). The different
    line types indicate different spectra.  The notes "NO" and "YES"
    indicate the spectra before and after the calibration of the sky
    respectively. The "Ref" means reference galaxy, MJ SUM is the
    final spectrum frame which is the sum of the single frames. The
    solid and open dots stand for the I band photometry profile 
    (see Sect. \ref{sec_dynres}) along
    the major and the minor axis, respectively.  In the top panel of
    the figure, the ratios $R$ of MJ3 to MJ1 (solid line) and MJ7
    (dotted line) are presented. \label{fig_skysub}}
\end{figure}

\begin{figure}
  \begin{center}
    \includegraphics[width=0.48\textwidth]{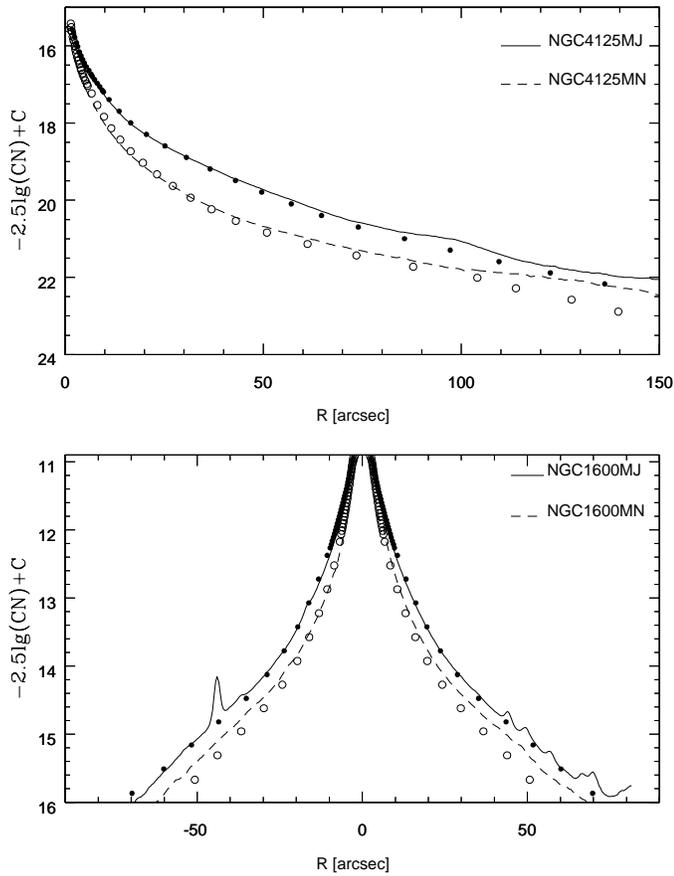}\
  \end{center}
  \caption{The comparison between the broad band surface brightness
    profiles (dots) and the ones derived from our summed spectra (lines) 
    for \object{NGC1600} and \object{NGC4125}. The full line and the filled dots are profiles 
    along the major axes. The  
    dashed line and the open dots show the profiles along the minor axes.
\label{fig_skysubproof}}
\end{figure}

The step of sky subtraction required particular care to minimize
systematic effects on the measured kinematics and line strengths in
the outer regions of our galaxies. We have to face the following
problems:  The HET queue scheduling operations and observing
conditions did not always allow to acquire blank sky spectra after the
science observations (see Table \ref{tab_log}). Moreover variability
in both the atmospheric absorption (due to different air masses and/or
non-photometric conditions) and sky level were present.  Finally, not
always a perfectly flat slit illumination was achieved.  In order to
account for these problems, we selected the spectra for each galaxy
where a sky spectrum with a uniform slit illumination was available, and
almost photometric conditions were achieved, yielding the largest galaxy
counts per pixel. 
To correct for the inhomogeneous slit illumination, we produced a 4th
to 6th order polynomial model of the sky spectra for each column in
spatial direction and subtracted it from the selected galaxy frames,
obtaining three reference science frames $G_r$ -- one per galaxy.
In the next step,
we calibrated the atmospheric absorption and sky level of the remaining
galaxy frames $G_i$ using the following procedure.  We computed
the fractional residuals between the scaled and the reference slit profiles
\begin{equation}
\label{eq_sky}
R(r)=|1-\frac{f_i^G \ast \langle G_r\rangle}{\langle (G_i\rangle -f_i^S\ast SKY_i)\rangle}|.
\end{equation}
and minimized it (see below).
Here $f_i^S$ is the scaling factor of the noise-free (i.e. the
polynomial model) sky frame taken after the galaxy frame $G_i$, when
available, or the average of the most uniform sky frames when not.
The symbol $\langle\rangle$ indicates the average in the wavelength
direction and $R(r)$ is a function of the position $r$ along the slit.
Moreover, $f_r^G$ is a scaling factor that takes into account the
different atmospheric transmissions.  We determined $f_i^G$ and $f_i^S$
iteratively such to minimize R, which in an ideal situation should be
zero at every radii.  Finally, we computed the resulting total galaxy
frame $G_{tot}$ as:
\begin{equation}
\label{eq_tot}
G_{tot}=G_r+\Sigma_i (G_i-f_i^S \ast SKY_i)/f_i^G.
\end{equation}
In practice, due to the non-uniformity of the slit illumination
function the function $R(r)$ is not always zero, but through the
summing in Eq. \ref{eq_tot} the differences should average out. We can
test the quality of the calibration by comparing the profile $\langle
G_{tot}\rangle$ with available broad band photometry. Figure
\ref{fig_skysub} shows some examples of the calibration procedure for
the spectra taken along the major axis of \object{NGC7619}.  The different type
lines indicate different slit spectra. The notes "NO" and "YES"
indicate spectra before and after calibration. The slit marked with
"Ref" is reference spectrum. The solid and open dots show the I broad
band photometry profile (see Sect. \ref{sec_dynres}) along major and
minor axis respectively. The top panel shows the ratios $R$ of MJ3 to
MJ1 (solid line) and MJ7 (dotted line). As it can be seen from the
figure, the MJ1 and MJ3 matched well with the photometry profiles from
the center to the outer parts. In contrast, the MJ7 profile has
larger deviations from the photometry profile in the outer regions
since the sky was over-subtracted on the left side and underestimated
on the right side. The summed spectrum agrees well with the broad band
photometry out to $r\approx 70$ arcsec.  The kinematics and line
strength profiles derived in Sect. \ref{sec_kinlin} extend out to
radii where the agreement with broad band photometry is always
good. Figure \ref{fig_skysubproof} shows the comparison between the
summed shifted slit profiles and the broad-band photometry for the
three galaxies.

The procedure described above removes the differences in the sky
continuum, but does not necessarily calibrate correctly the sky
emission lines. Therefore, following \citet{2009arXiv0910.5590S} we
constructed a separate zero continuum sky emission line spectrum that
was scaled and subtracted from the summed spectra.  As a final step,
the continuum was removed by fitting and dividing by a fourth order
polynomial along the wavelength direction.

Finally, in order to gauge the residual systematic errors, we repeated
the extraction of the kinematics increasing or decreasing the scale
factor for the sky level by the value of the rms difference in the
normalized slit illumination function (see upper plot of
Fig. \ref{fig_skysub}).  The effect on the extracted kinematics is
most significant in the case of \object{NGC1600} and is indicated by the dotted
errors in Fig.~\ref{fig_kin}.

\section{Kinematics and line indices}
\label{sec_kinlin}

\subsection{Kinematic profiles}
\label{sec_kin}

The line-of-sight velocity distributions (LOSVDs) and kinematic
parameters were determined from the continuum-removed spectra,
rebinned radially to obtain almost constant signal-to-noise ratio,
using the Fourier Correlation Quotient (FCQ) method
\citep{1990A&A...229..441B} with the implementation described in
\citet{2009arXiv0910.5590S} that allows for the presence of emission
lines.  In order to minimize the mismatching, we use the templates
from a subset of the stellar spectra library of
\citet{1999ApJ...513..224V}. This library contains about a thousand
synthetic single-stellar-population spectra covering the wavelength
range from 4800 to 5470 {\AA} with the resolution of 1.8 {\AA}. We
used the library with the age of 1.00 to 17.78 Gyr and the
metallicities from -1.68 to 0.2. We first set all of the library
spectra to the resolution of our galaxy spectra and find the best
fitting template for each radial bin according to the lowest RMS value
of the residual (reaching typically  1\% of the initial flux). 
If emission lines are detected, Gaussians are fitted
to the residuals above the best-fit template and subtracted from the
galaxy spectrum to derive cleaned spectra. The kinematic fit is then
redone using these cleaned spectra.  We detect emission only in the
spectra of \object{NGC4125}.

The stellar kinematic profiles of both the major and the minor axes of
\object{NGC1600}, \object{NGC4125} and \object{NGC7619} are shown in Fig.~\ref{fig_kin}, where we
show the rotational velocity, the velocity dispersion and the
Gauss-Hermite parameters $H_3$ and $H_4$. The curves are folded with
respect to the center of the galaxies, filled and open symbols stand
for different sides\footnote[1]{For \object{NGC4125} and 7619: The filled symbols
  show the kinematic profiles on south-west (SW) (receding) side along
  the major axis and north-west (NW) (approaching) side along the 
  minor axis. For \object{NGC1600}:
  The open symbols show the kinematic profiles on the
  north-east (NE) approaching side along the major axis and south-east 
(SE) side along 
the minor axis.} of the galaxies. The $a_e = R_e \cdot \epsilon^{-1/2}$ and
$b_e = R_e \cdot \epsilon^{1/2}$ are labeled, where the R$_e$ is the
effective radius and the $\epsilon$ is the apparent axial ratio.  The
long dashed lines show the instrument dispersion. The solid lines
present the kinematics from the dynamic models (see Sect.
\ref{sec_dynres}). Table \ref{tab_stellardata} gives format examples
of the measured stellar kinematics with statistical and systematic
errors. The full listing is available electronically.

Also the previous studies of kinematics are shown in the plot. The
circles present the data of this work, the squares show the previous
work of \citet{1994MNRAS.269..785B} for \object{NGC1600} and \object{NGC4125},
\citet{1995ApJ...438..539F} for \object{NGC7619}. As it can be seen from the
figures, the agreement is generally good except for the major axis
data of \object{NGC1600}. In all cases our profiles extend to larger radii and
show less scatter than those of previous investigation. \object{NGC4125} has
dust and ionized gas aligned along the major axis
\citep{1984AJ.....89..356B,1989ApJ...346..653K,1995A&A...297..643W}.
We also measured the kinematics of ionized gas by fitting Gaussians to
the emission lines of H$\beta$, [OIII]$\lambda\lambda$ 4958,5007 and
[NI]$\lambda\lambda$ 5197,5200.  The kinematics of the gas are shown
in Fig. \ref{fig_gas}. Table \ref{tab_gasdata} gives format examples
of the measured gas kinematics. The full listing is available
electronically.

\begin{figure*}
\begin{center}
\begin{tabular}{c}
\includegraphics[width=0.49\textwidth]{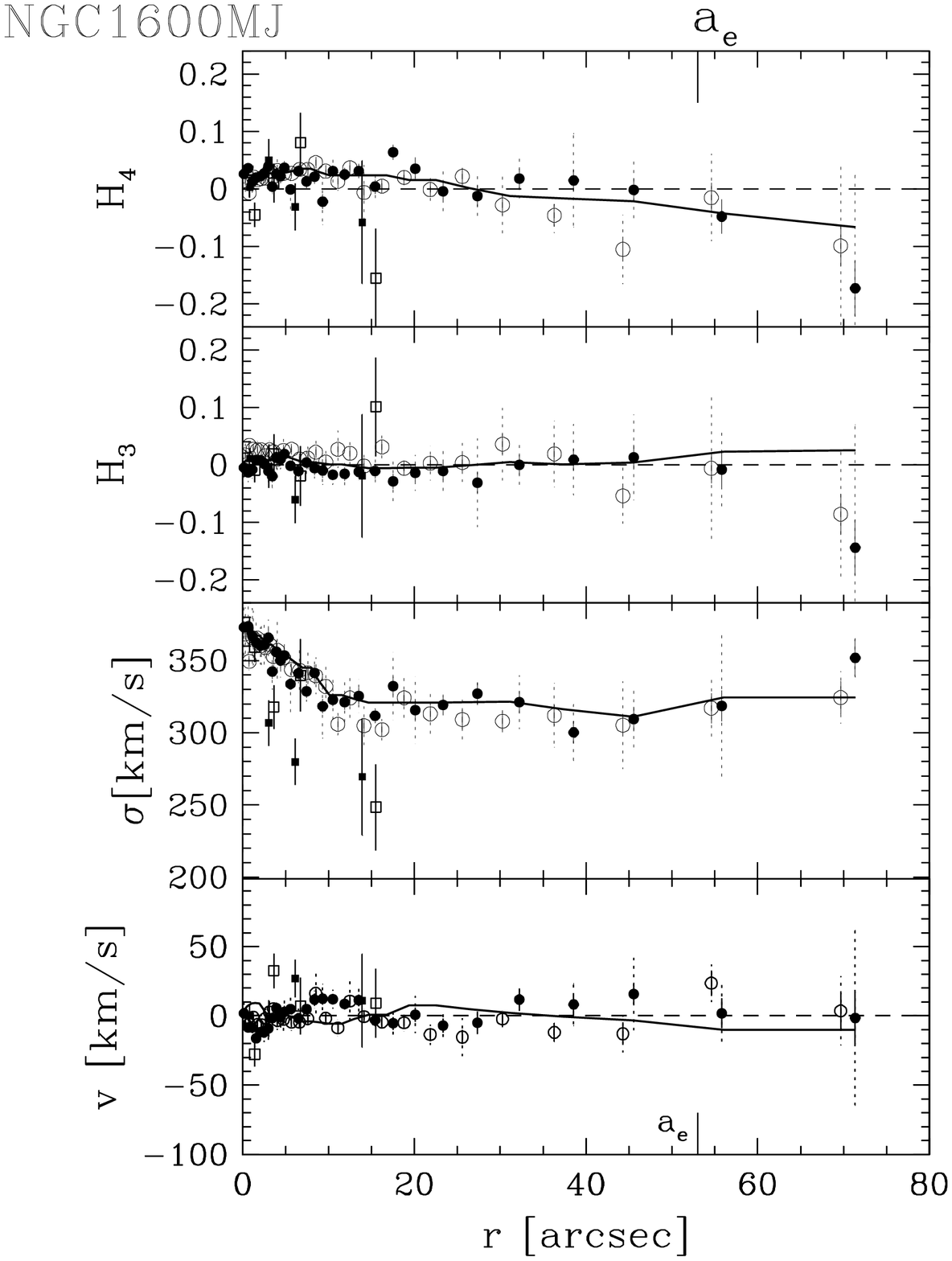}
\includegraphics[width=0.48\textwidth]{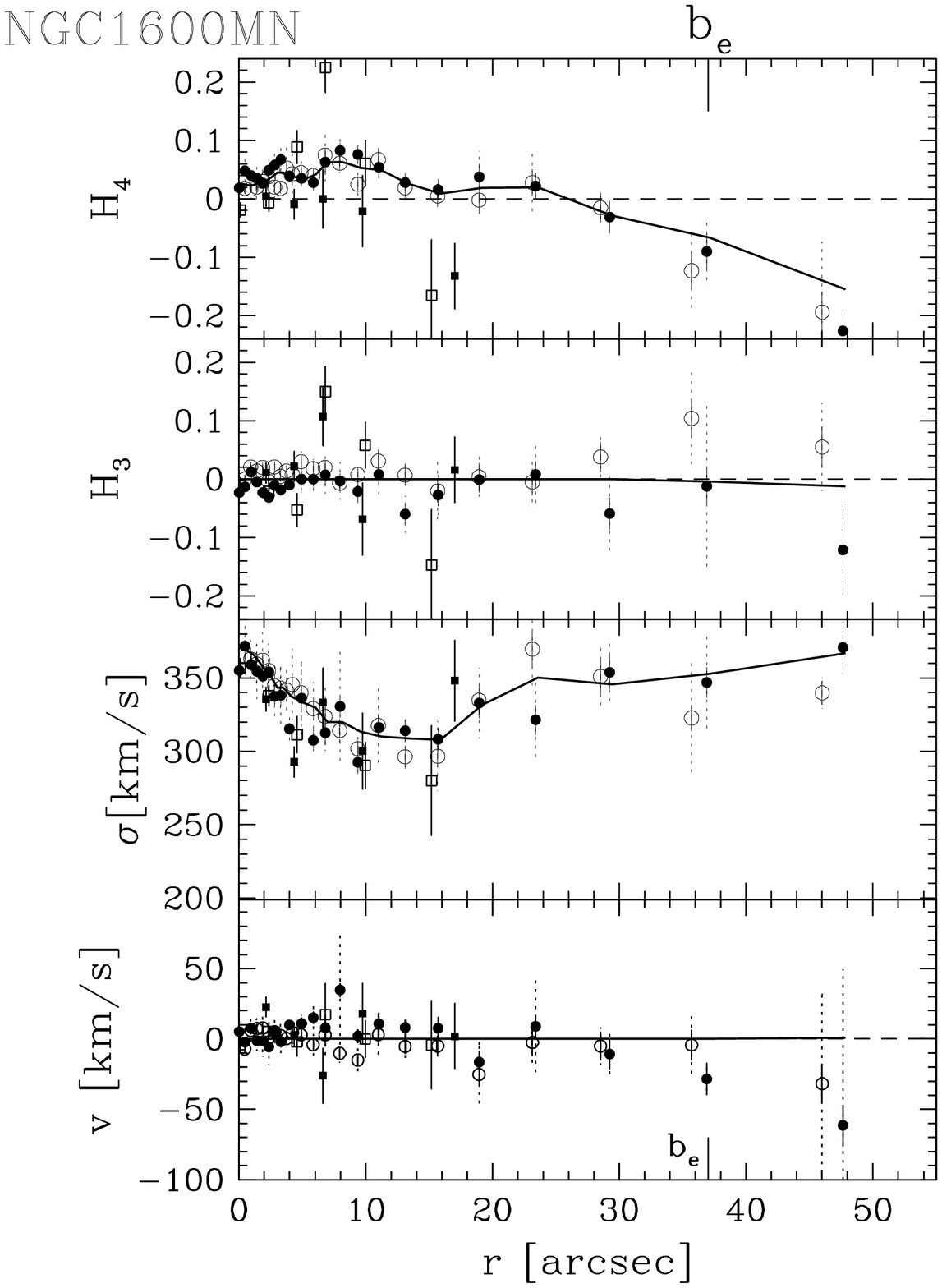}
\end{tabular}
\end{center}
\caption{Kinematic parameters along the major (left) and the
  minor (right) axis of our galaxies, the corresponding names
  are labeled.  From bottom to top in each panel we show: (1)
  Rotation velocity, (2) Velocity dispersion, (3) and (4)
  Gauss-Hermite parameters $H_3$ and $H_4$. The solid lines present the
  kinematic moments from dynamic models (see Sect. \ref{sec_dynres}).
  The doted lines show the systematic uncertainty due to sky subtraction.
  The curves are folded with respect to the nucleus of the
  galaxies, filled and open symbols stand for different sides
  of the galaxies.  The squares show the data published in
  \citet{1994MNRAS.269..785B}. \label{fig_kin}}
\end{figure*}

\addtocounter{figure}{-1}
 \begin{figure*}
 \begin{center}
 \begin{tabular}{c}
 \includegraphics[width=0.48\textwidth]{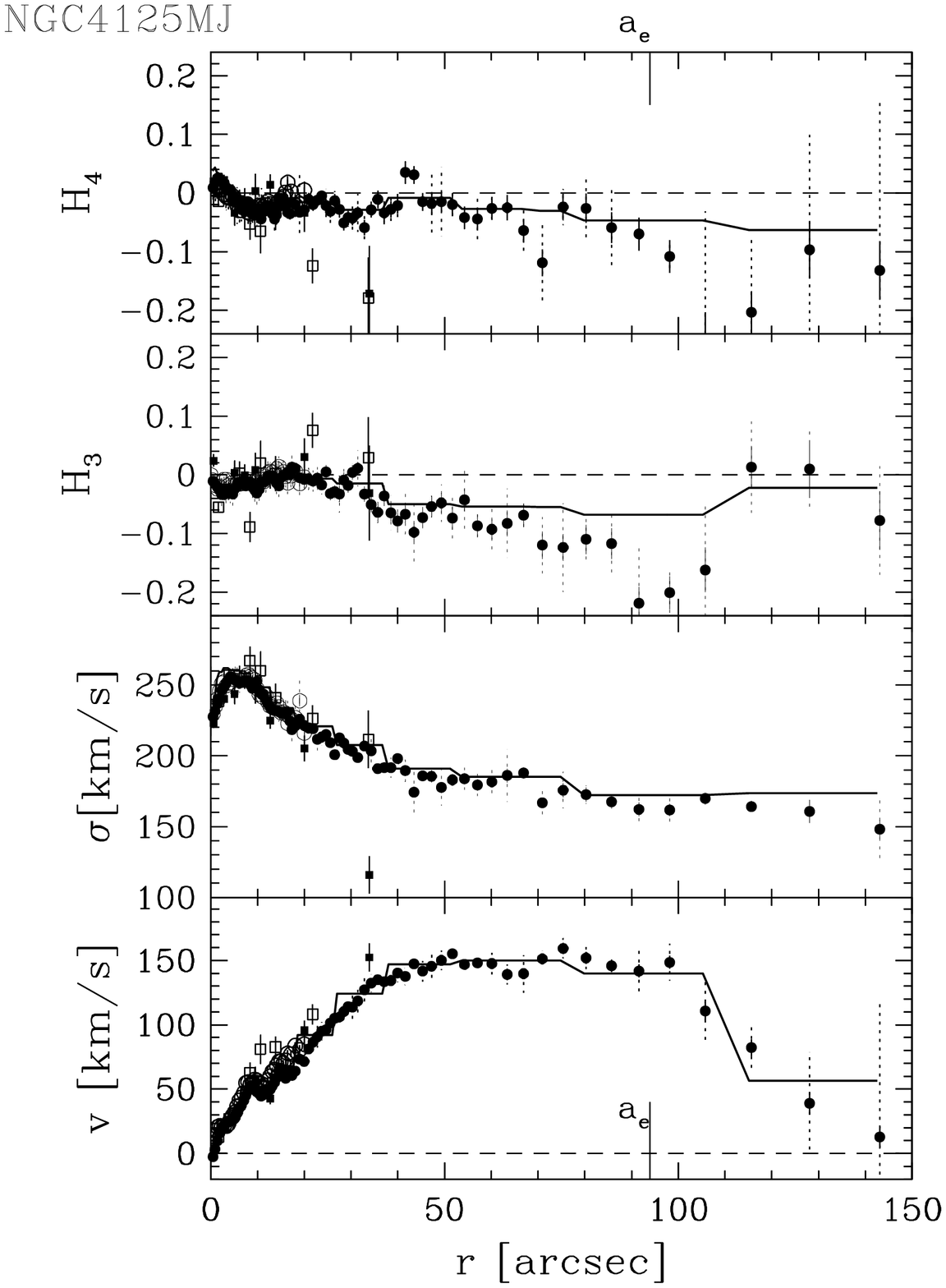}
 \includegraphics[width=0.467\textwidth]{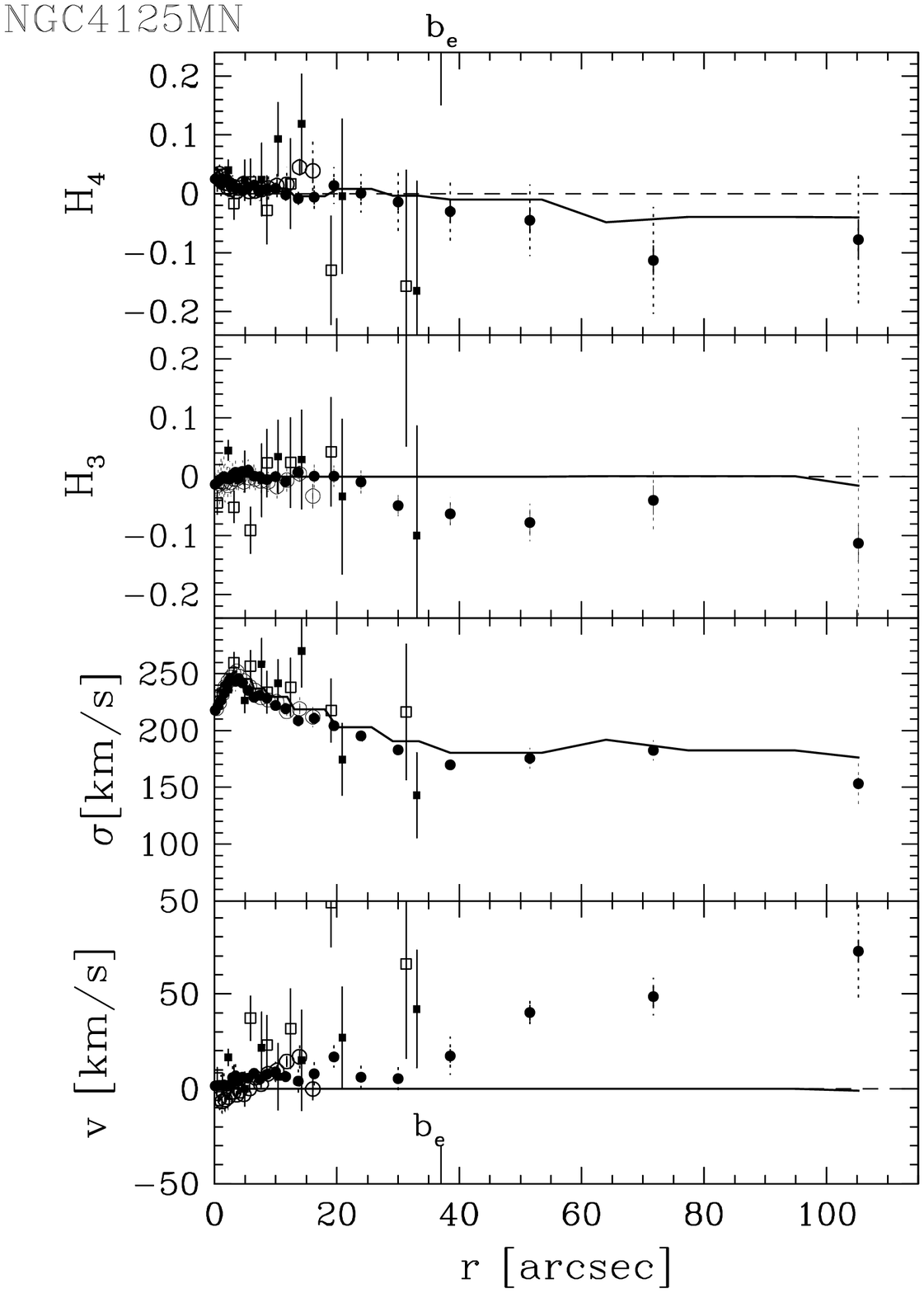}
 \end{tabular}
 \end{center}
 \caption{Continued. The stellar kinematics  of
   \object{NGC4125}.  The squares show the data published in
   \citet{1994MNRAS.269..785B}. The solid lines present the
   kinematics from dynamic models (see Sect. \ref{sec_dynres}).}
 \end{figure*}

 \addtocounter{figure}{-1}
\begin{figure*}
 \begin{center}
 \begin{tabular}{c}
 \includegraphics[width=0.48\textwidth]{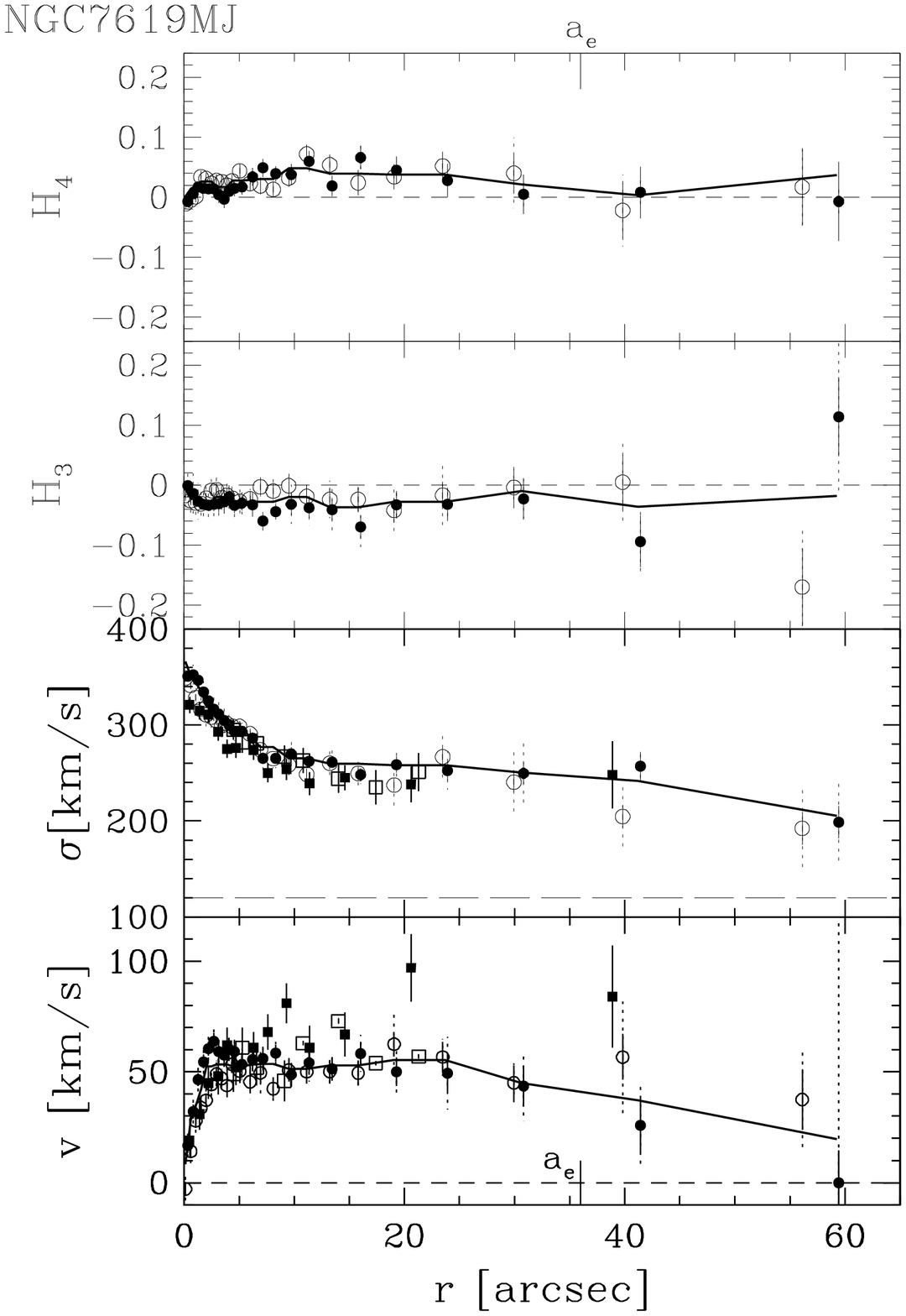}
 \includegraphics[width=0.485\textwidth]{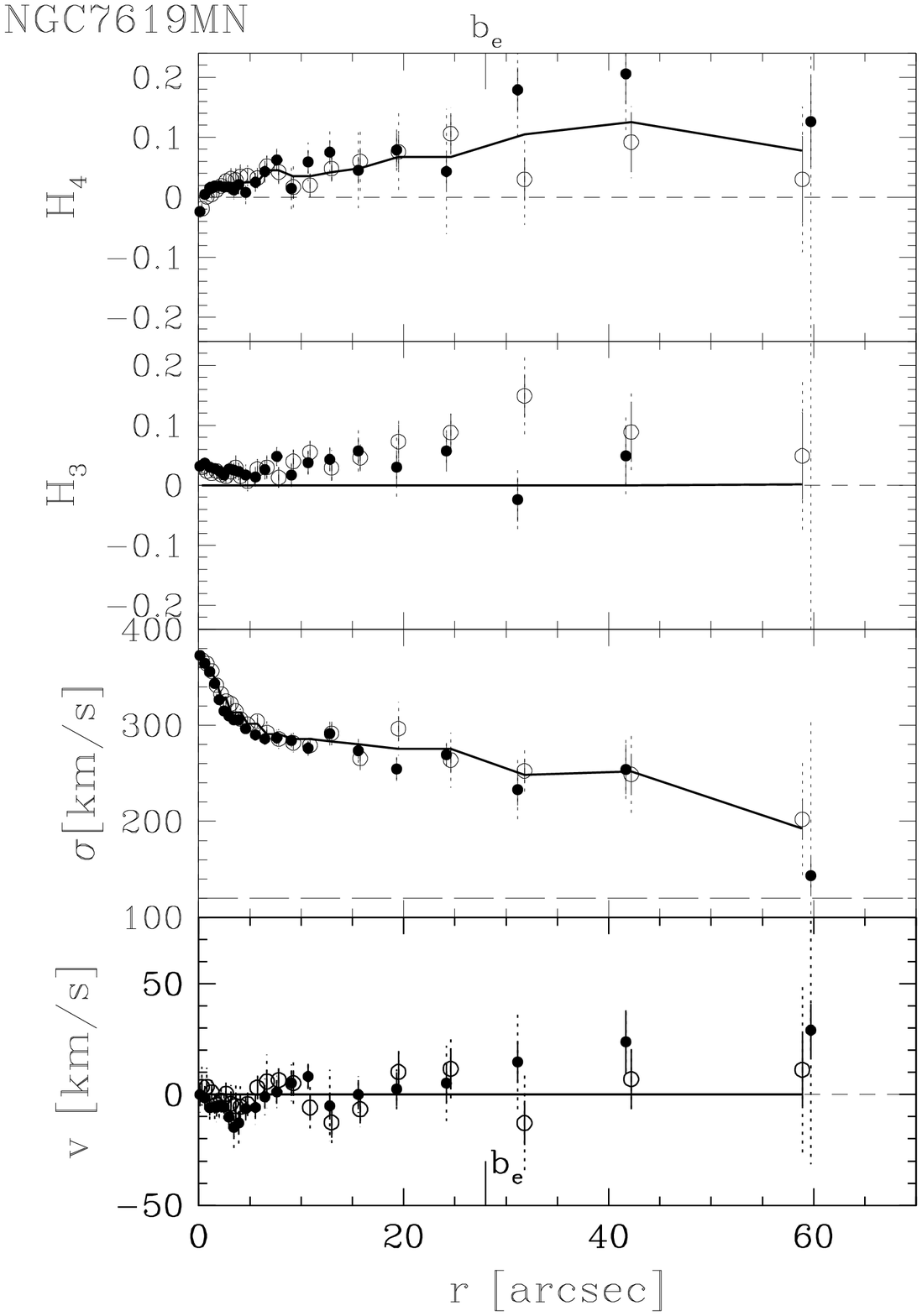}
 \end{tabular}
 \end{center}
\caption{Continued. The stellar kinematics along
  the major and the minor axes of \object{NGC7619} is shown; the squares present 
  the measurement of \citet{1995ApJ...448..119F}. The solid lines present the
  kinematics of the dynamic models (see details in Sect. 
\ref{sec_dynres}). 
The long dashed lines indicate the instrumental dispersion. \label{fig_kin7619}}
 \end{figure*}
\begin{table*}
\caption{Format example of the measured stellar kinematics as a function of distance from the center (positive: east, negative: 
west) for the different position 
angles. The full table is available electronically.} 
\label{tab_stellardata}
\begin{tabular}{ccccccc}
\hline
Galaxy & R & PA & $V_{stars}\pm dV_{stat}\pm dV_{sys}$ & $\sigma_{stars}$ & $H_3$ & $H_4$ \\ 
& (\arcsec) & (deg) & (km/s) & (km/s) & & \\
\hline
 \object{NGC1600} & -69.67 & 9 & $-3.8\pm12.6\pm24.9$ & $324.4\pm11.3\pm17.9$ & $0.09 \pm 0.04 \pm 0.11$ & $-0.10\pm0.04\pm0.15$\\
 \hline
\end{tabular}
\end{table*}

\begin{table*}
\caption{Format example of the measured gas kinematics for \object{NGC4125} as a function of distance from the center (positive: east, negative: 
west) for the different position 
angles. The full table is available electronically.} 
\label{tab_gasdata}
\begin{tabular}{ccccccccc}
\hline
Galaxy & R & PA & $V_{gas}$ & $\sigma_{gas}$ & H$\beta$ & [OIII]/H$\beta$ & [NI]/H$\beta$ \\ 
       & (\arcsec) & (deg) & (km/s) & (km/s) & \AA & & \\
\hline
\object{NGC4125} & -81.3 & 82	& $-3.4\pm25.5$ & $180.9\pm25.6$ & $0.06\pm0.02$ & $6.0\pm1.6$&$1.7\pm0.03$\\
\hline
\end{tabular}
\end{table*}
\subsection{Line strength indices}
\label{sec_lin}

\begin{figure}
   \centering
   \includegraphics[width=0.48\textwidth]{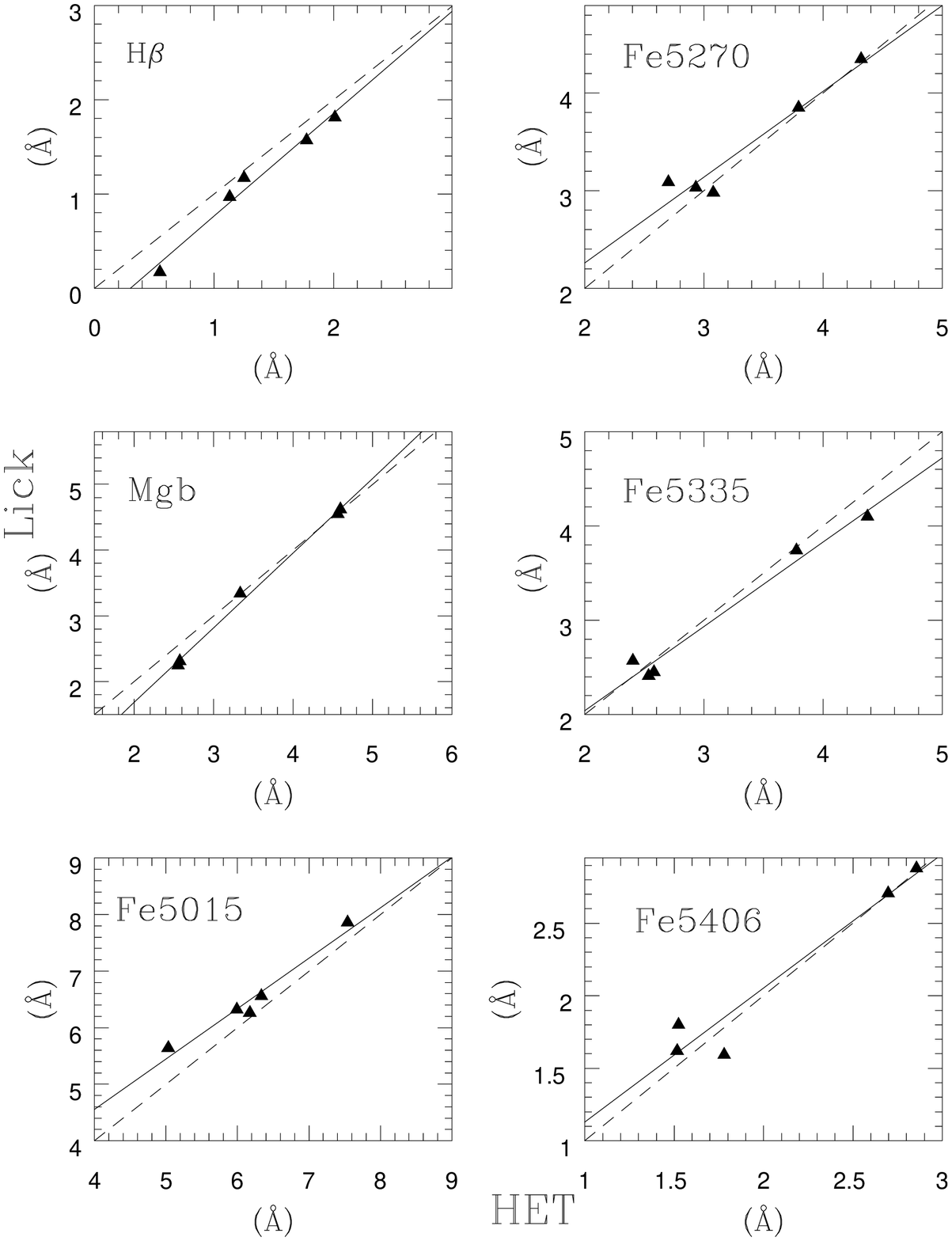}
   \caption{The comparisons between the values for the lines strength
     indices derived in this work and the standard Lick/IDS indices.
     The dashed line presents the 1:1 relation, the solid lines
     indicate the linear relation fit to the data. \label{fig_lick}}
\end{figure}

Line-strength indices were defined by
\citet{1984ApJ...287..586B,1994ApJS...94..687W} and redefined by
\citet{1998ApJS..116....1T}. In this work, the index windows follow
the definition of \citet{1998ApJS..116....1T}. For all of our
galaxies, six line strength indices have been measured from the
cleaned spectra along the major and minor axis. Before measuring the
indices, our spectra were degraded to the resolution of Lick/IDS
systems. We then corrected the indices for the velocity dispersion
using template stars and the value for $\sigma$ derived in the
previous section. Finally, the observational data need to be corrected
to the Lick/IDS system.  To do this we observed 5 stars from the
Lick/IDS library using the same instrumental configuration used for
the science objects and derived the offsets between our data and the
Lick/IDS system.  The comparison of our data with the Lick standard
systems can be found in Fig. \ref{fig_lick}. The dashed line presents
the 1:1 relation, the solid lines indicate the linear fit the
data. The data are in good agreement with the Lick/IDS systems, as
found in \citet{2009arXiv0910.5590S} using a larger set of Lick
standards observed with LRS and HET at a better resolution. Only for
the $H\beta$ there is a small, marginally significant offset, possibly
related to a defect column in the CCD \citep[see discussion
in][]{2009arXiv0910.5590S}. The linear relation between indices in
LICK/IDS systems and our measurements are listed in Table
\ref{tab_lick}. We conclude that the deviation between our
measurements and the Lick system can be ignored, but we take into
account the RMS of the calibration lines into the final error budget,
by adding it in quadrature to the statistical error of each index.

  \begin{table}
    \caption{The linear transformation between indices measured by us and 
      in LICK systems. The second column displays the best linear relations, 
      and the last column show the ratio of LICK systems to our measurements. 
\label{tab_lick}}
  \begin{center}
  \begin{tabular}{lccl}
  \hline
  Index   & Linear relation &RMS &Lick/HET \\
            &(Lick=a$\times$HET+b) & &\\
  \hline
  $H\beta$  &1.091($\pm$ 0.096)-0.327($\pm$ 0.134) &0.085  &0.780 $\pm$ 0.117\\
  $Mg_b$    &1.137($\pm$ 0.025)-0.549($\pm$ 0.129) &0.073  &0.967 $\pm$ 0.0245\\
  $Fe5015$  &0.892($\pm$ 0.105)+0.980($\pm$ 0.663) &0.147  &1.052 $\pm$ 0.0276\\
  $Fe5270$  &0.879($\pm$ 0.135)+0.499($\pm$ 0.471) &0.143  &1.037 $\pm$ 0.0291\\
  $Fe5335$  &0.903($\pm$ 0.085)+0.255($\pm$ 0.321) &0.117  &0.990 $\pm$ 0.0249\\
  $Fe5406$  &0.974($\pm$ 0.127)+0.171($\pm$ 0.265) &0.142  &1.066 $\pm$ 0.0373\\
 \hline
 \end{tabular}
 \end{center}
 \end{table}

 As in \citet{2009arXiv0910.5590S}, we do not consider the molecular
 indices $Mg_1$ and $Mg_2$ that are affect by inaccurate spectral
 flatfielding towards the end of the slit, where vignetting becomes
 important.  Figure \ref{fig_lin} shows the lines strength indices
 profiles along the major (left) and minor (right) axis of our three
 galaxies. The name and position (major axis and minor axis) are
 labeled in the figure, the asterisks show the measured line strengths
 as the function of the radii in arcseconds along the major and minor
 axes, the solid lines show the model predictions (see Sect.
 \ref{sec_stelpop}). Finally, as for the kinematics, we indicate the
 systematic variations due to sky subtraction with dotted errors. The
 blue symbols show values from the literature (see below). The agree
 well with our measurements.  Table \ref{tab_Lickdata} gives format
 examples of the measured Lick indices with statistical and systematic errors 
as a function of distance and
 position angle, respectively.  The full listing is available
 electronically.

\begin{figure*}
 \begin{center}
 \begin{tabular}{c}
 \includegraphics[scale=0.4]{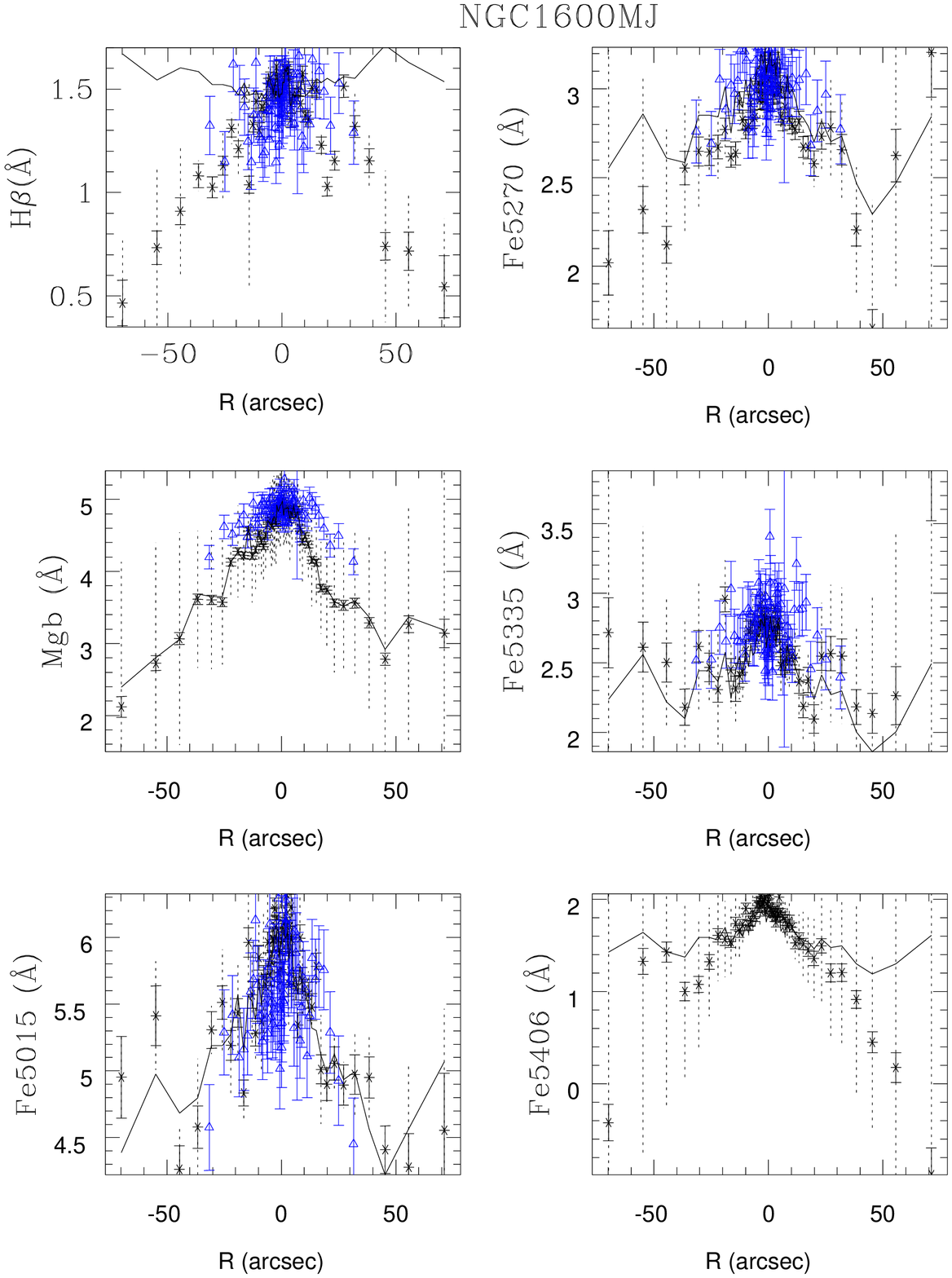}
 \includegraphics[scale=0.4]{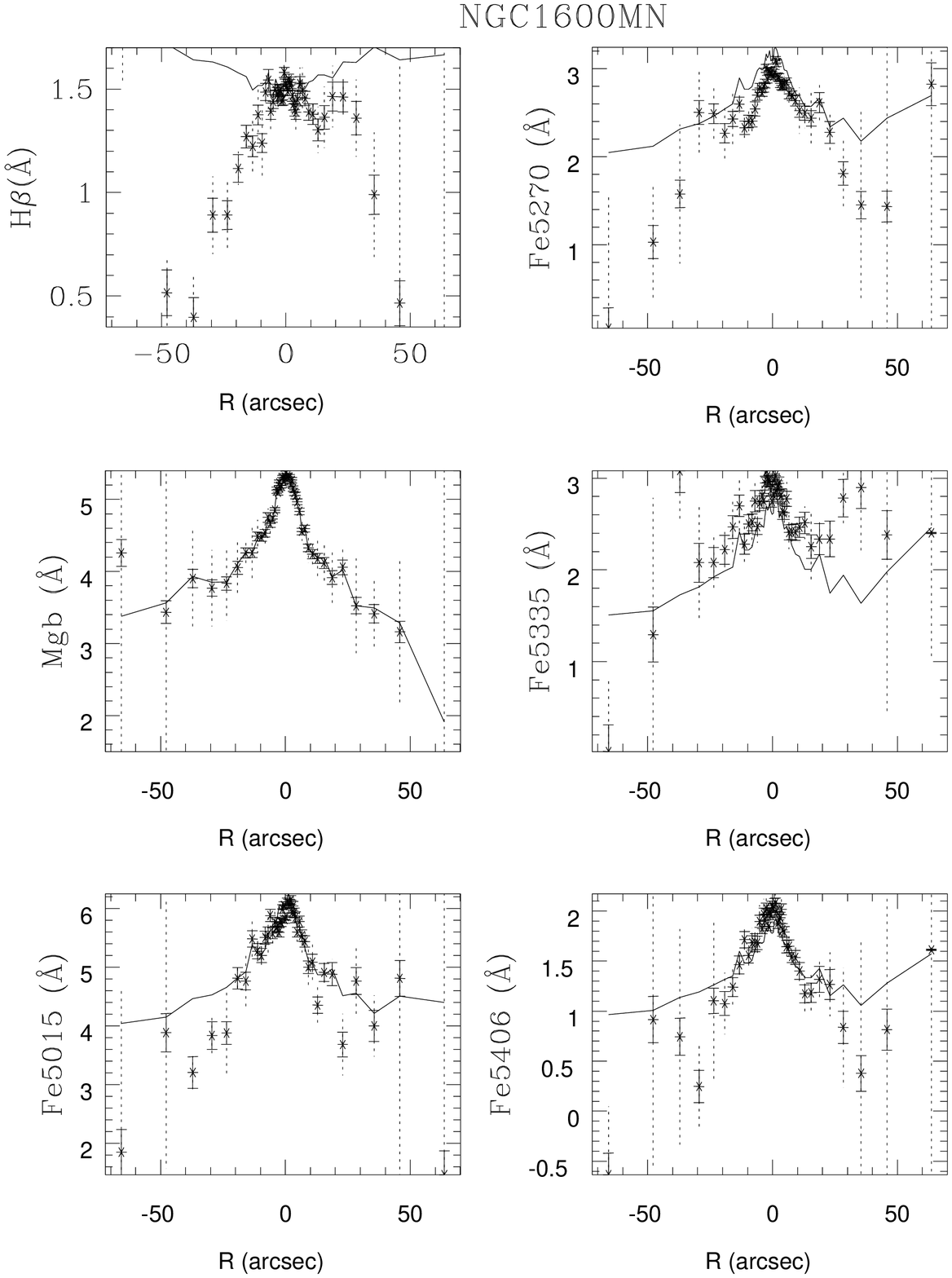}
 \end{tabular}
 \end{center}
 \caption{The lines-strength indices along the major (left) and minor
   axis (right). The name of the indices are labeled
   and the galaxies' names are also noted.
   The dotted lines show the size of the systematic
   errors due to sky subtraction. The solid lines show the
   models (TMB03) predicted lines strength profiles along the axes.
   The blue open triangles are measurement of \citet{2007MNRAS.377..759S}.
   \label{fig_lin}}

 \end{figure*}

\addtocounter{figure}{-1}
 \begin{figure*}
 \begin{center}
 \begin{tabular}{c}
 \includegraphics[scale=0.4]{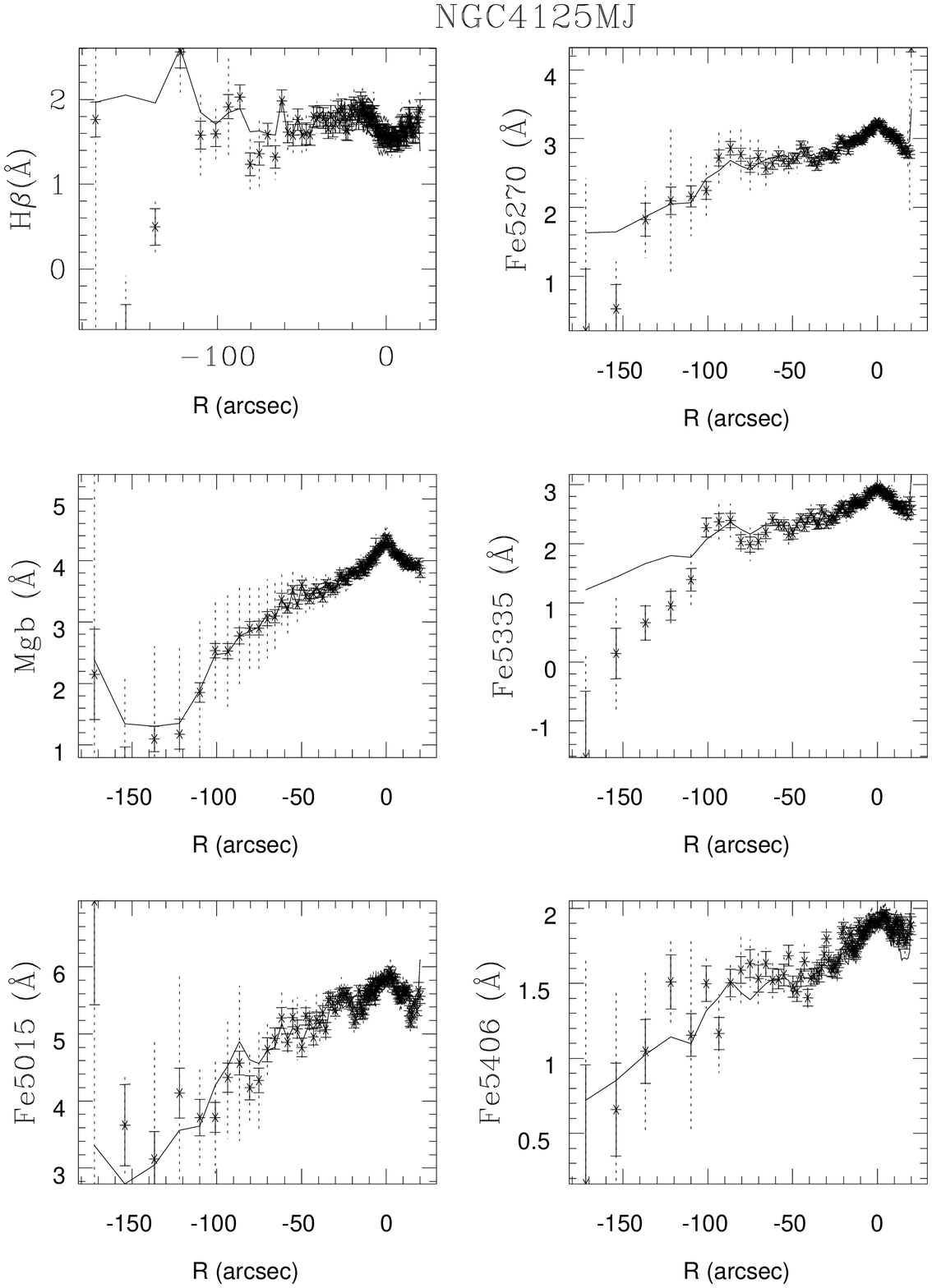}
 \includegraphics[scale=0.4]{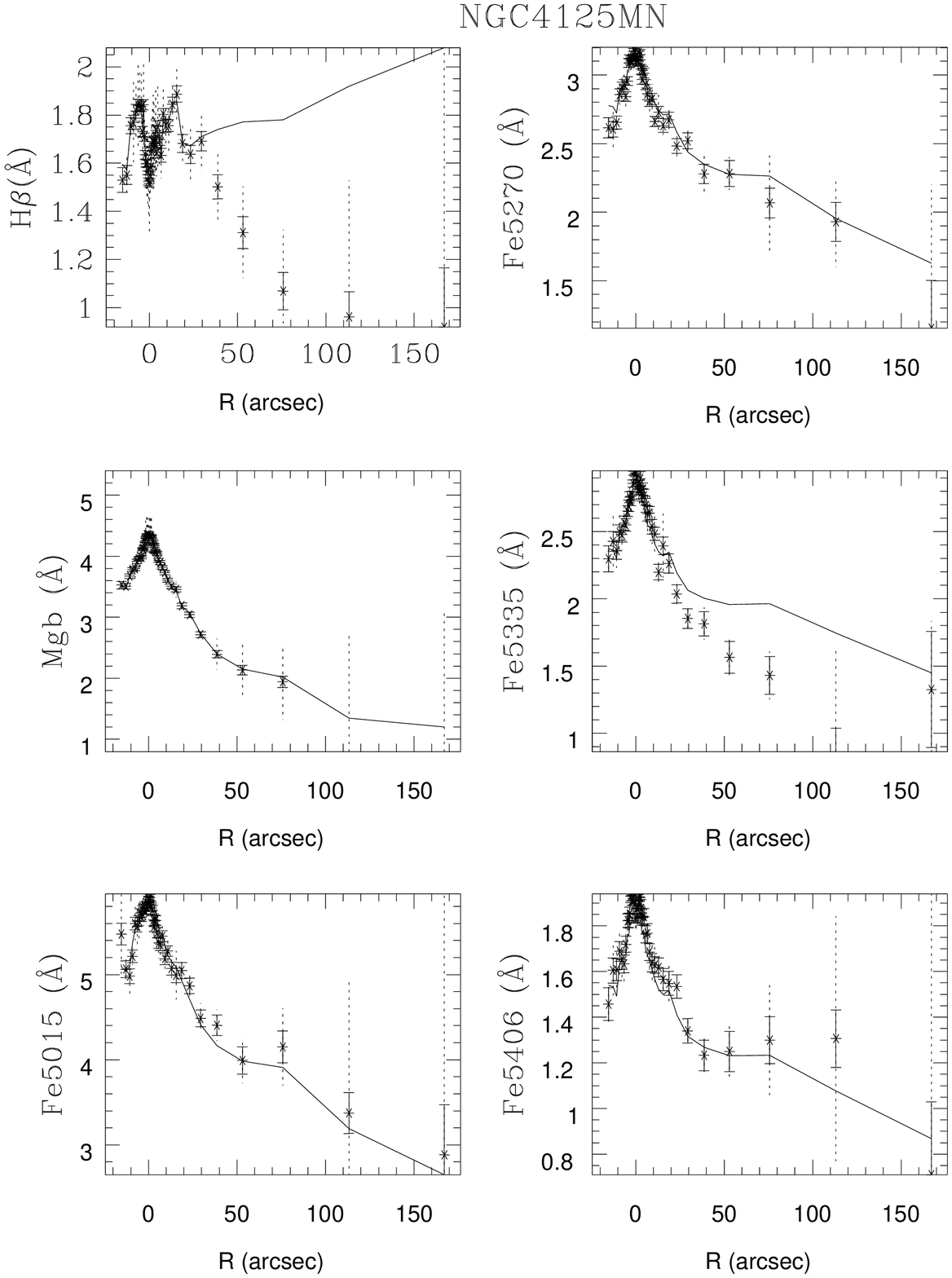}
 \end{tabular}
 \end{center}
 \caption{Continued.}
 \end{figure*}

\addtocounter{figure}{-1}
 \begin{figure*}
 \begin{center}
 \begin{tabular}{c}
 \includegraphics[scale=0.4]{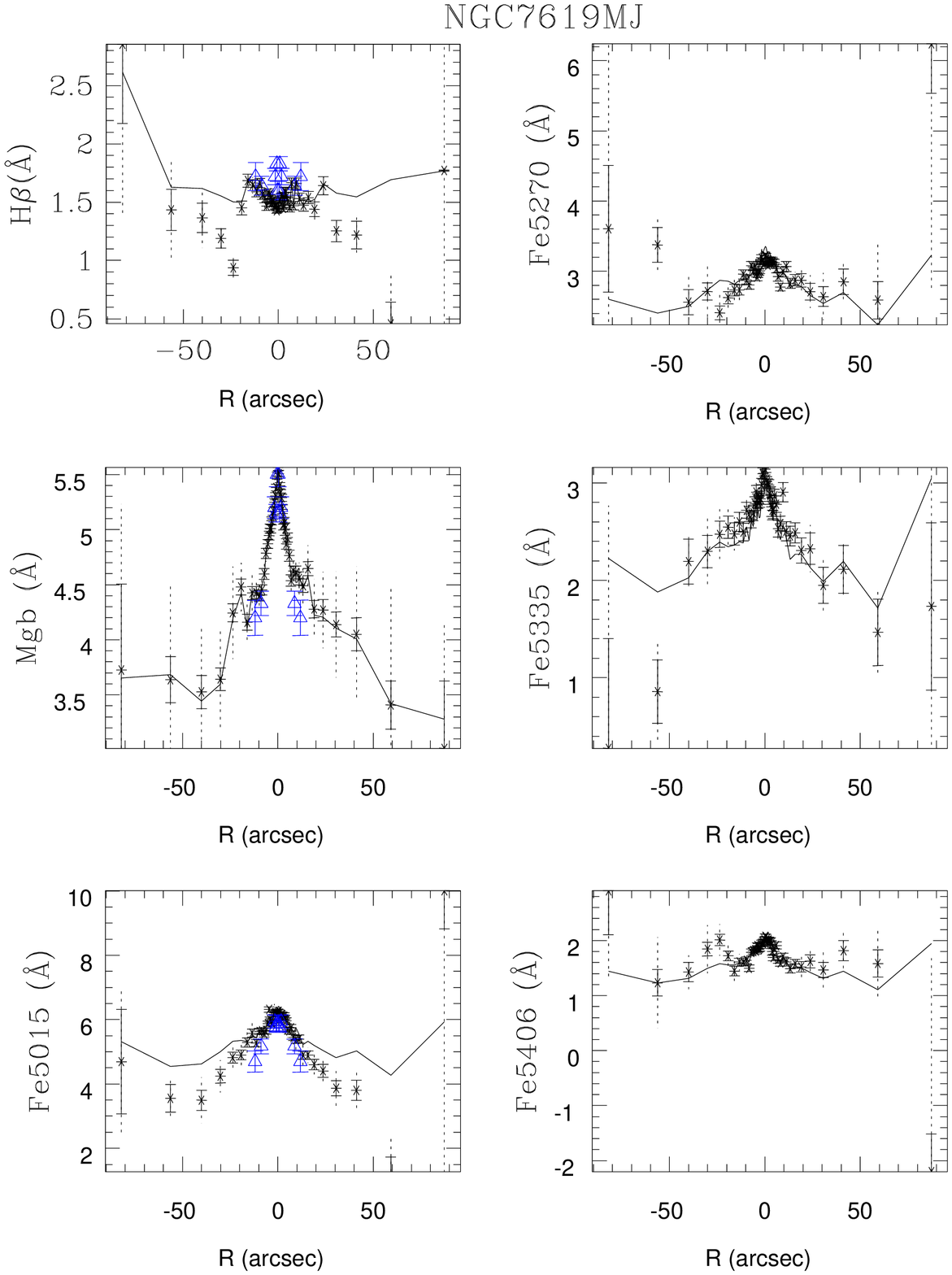}
 \includegraphics[scale=0.4]{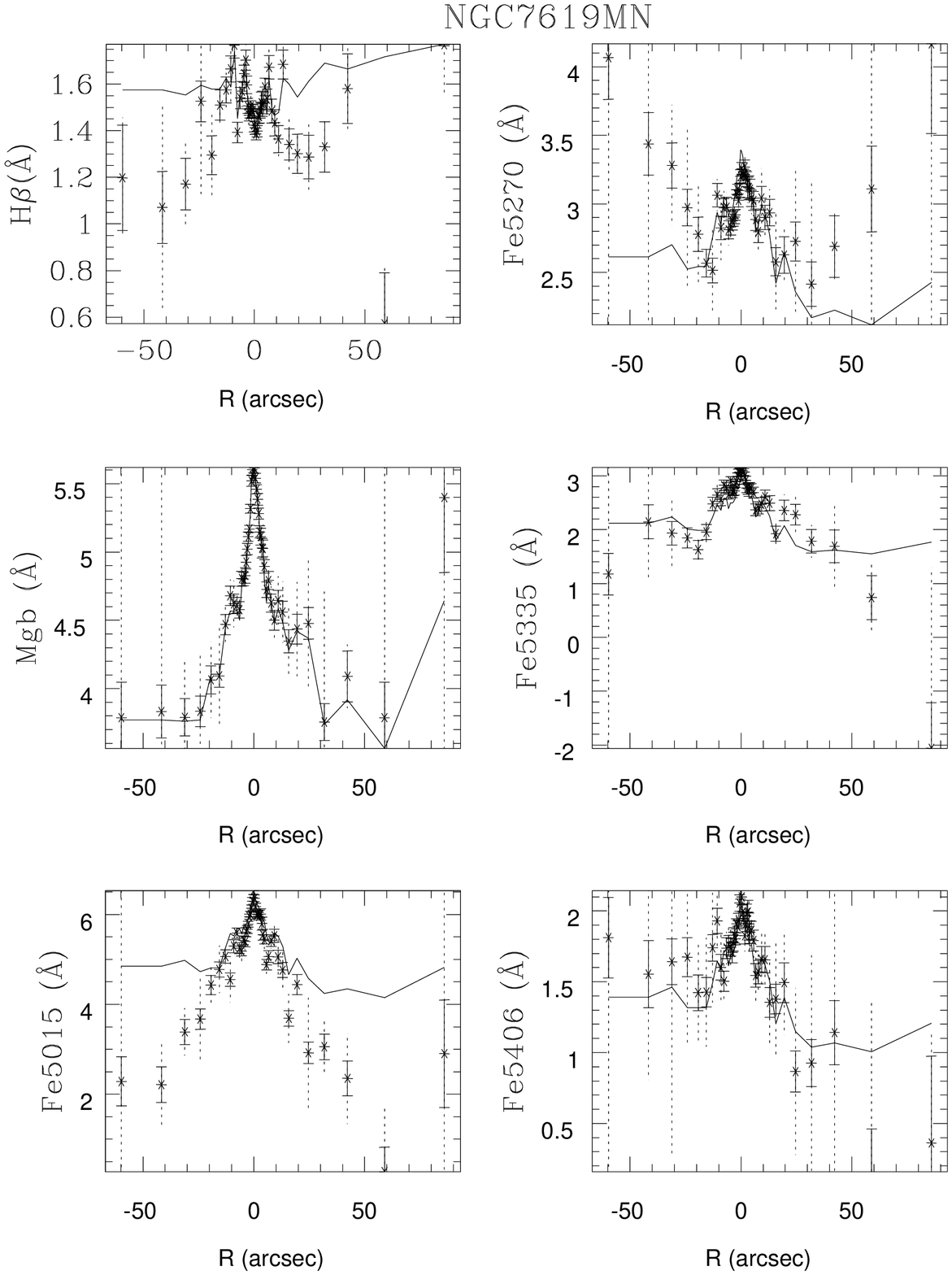}
 \end{tabular}
 \end{center}
 \caption{Continued. The blue open triangles showing the data taken from \citet{1995ApJ...448..119F}.}
\end{figure*}

\begin{table*}
\caption{Format example of the measured Lick indices as a function of distance from the center (positive: east, negative: 
west) for the different position 
angles. The full table is available electronically.} 
\label{tab_Lickdata}
\begin{tabular}{ccccccccc}
\hline
Galaxy & R & PA & H$\beta\pm d$H$\beta_{stat}\pm d$H$\beta_{sys}$ & Mgb & Fe5015 & Fe5270 & Fe5335 & Fe5406 \\
 & (\arcsec) & (deg) & (\AA) & (\AA) & (\AA) &(\AA) &(\AA) &(\AA) \\
\hline
\object{NGC1600} & -69.9 & 9 &  $0.5\pm0.1\pm0.3$ & $2.1\pm0.1\pm2$ & $4.9\pm0.3\pm0.3$ & $2.0\pm0.2\pm1.2$ & $2.7\pm0.3\pm1.5$ & $-0.4\pm0.2\pm3.1$\\       
\hline
\end{tabular}
\end{table*}

\subsection{Comments on individual galaxies}
\label{sec_comments}

\emph{\object{NGC1600}:} This is the most distant galaxy of the sample and it is
classified as E3.  Our kinematic data extend to 70~arcsec from
the center, almost about four times the radius of
\citet{1994MNRAS.269..785B}.  The kinematic profiles show no
rotation along both axes and high velocity dispersion. The
velocity dispersion shows a steep gradient inside 15~arcsec, 
becomes flat along the major axis and even increases along the
minor axis outwards.  The galaxy shows a weakly
asymmetric profile of Mgb, Fe5015 and Fe5335 in the outer
part.  Moreover, it has low
H$\beta$ line strength, particularly in the outer parts.  
Because accurate measurements of H$\beta$
always suffer from the contamination of emission lines or sky
lines, we carefully checked the spectra, and no emission lines
or sky lines were detected. 
Our instrumental resolution could be too low to detect the weak emission
lines embed in the bandpass or some weak absorption features
of other elements, such as CrI $\lambda$4885 {\AA}, FeI
$\lambda$4891 {\AA}, set in the pseudo-continuum of H$\beta$
definition \citep{2005A&A...438..685K, 2005A&A...439..997P}.

\emph{\object{NGC4125}:} This galaxy shows strong ionized gas emission along
the major axis. The stellar kinematics were measured out to 1.7~a$_e$
on SW side along the major axis and 3.5~b$_e$ on NW side along the minor axis.
 As Fig. \ref{fig_kin} shows, the kinematic profiles are fairly symmetric,
and show a strong velocity gradient along the major axis; inside
50~arcsec the rotational velocity increases rapidly from 0 to
150~km/s, remains remains nearly flat out to 100 arcsec, and declines
at larger radii.

Furthermore, it is clear from Fig. \ref{fig_kin} that the \object{NGC4125}
shows rotation along the minor axis which could indicate triaxiality
\citep{1984AJ.....89..356B}, or be the sign of unsettled material left
over from a past merger and visible on images of the galaxy. The
velocity dispersion profile of the stars shows a depression in the
inner 8~arcsec, on both the major and the minor axes. This depression
suggests that there is a colder substructure in the innermost center.
The galaxy presents positive gradients of H$\beta$ index both along
major axis and minor axis in the center, while it remains flat towards
outer part. The H$\beta$ profile along major axis exhibits weak
asymmetry.

\begin{figure*}
\begin{center}
\begin{tabular}{c}
\includegraphics[scale=0.45]{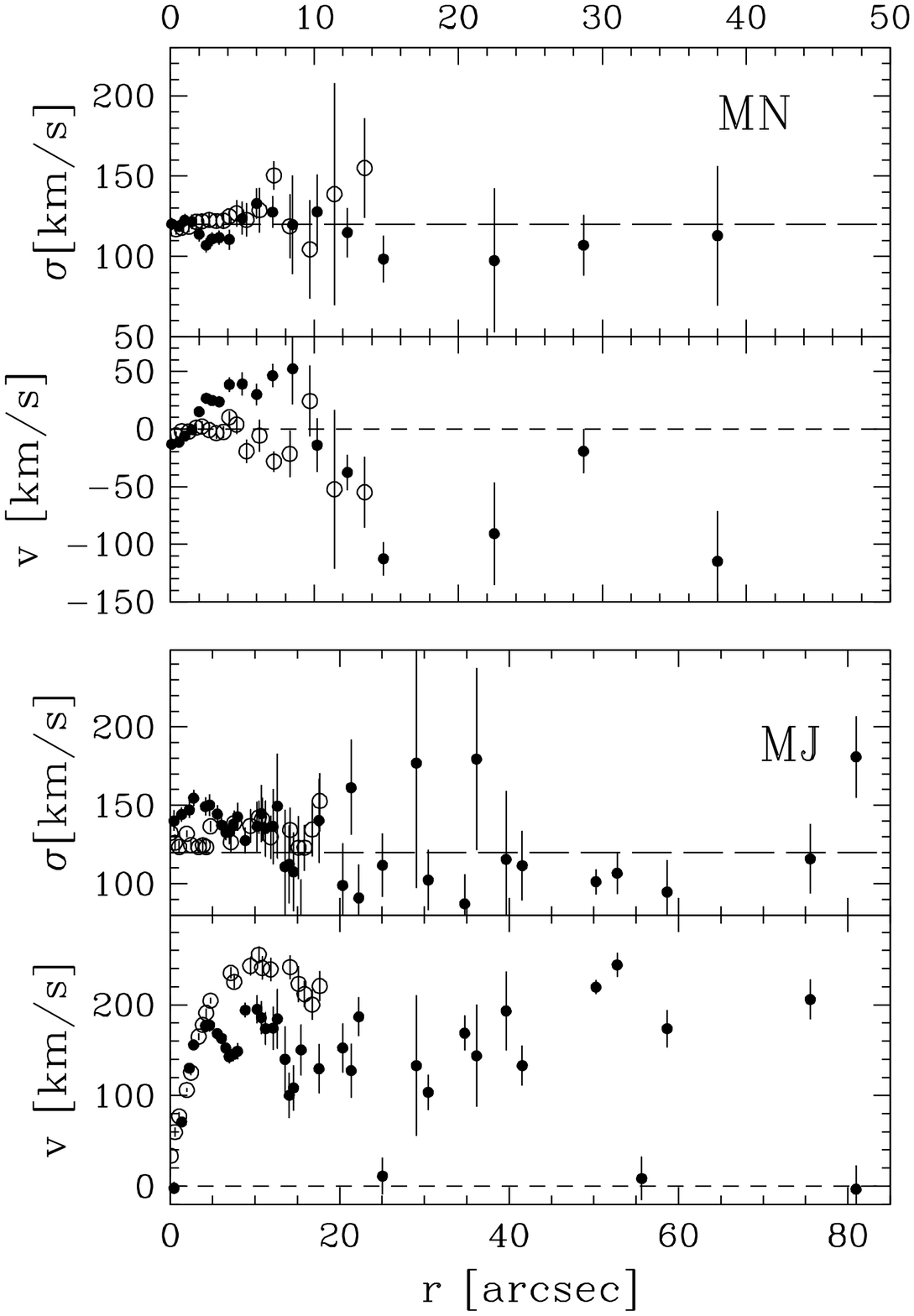}
\includegraphics[scale=0.4]{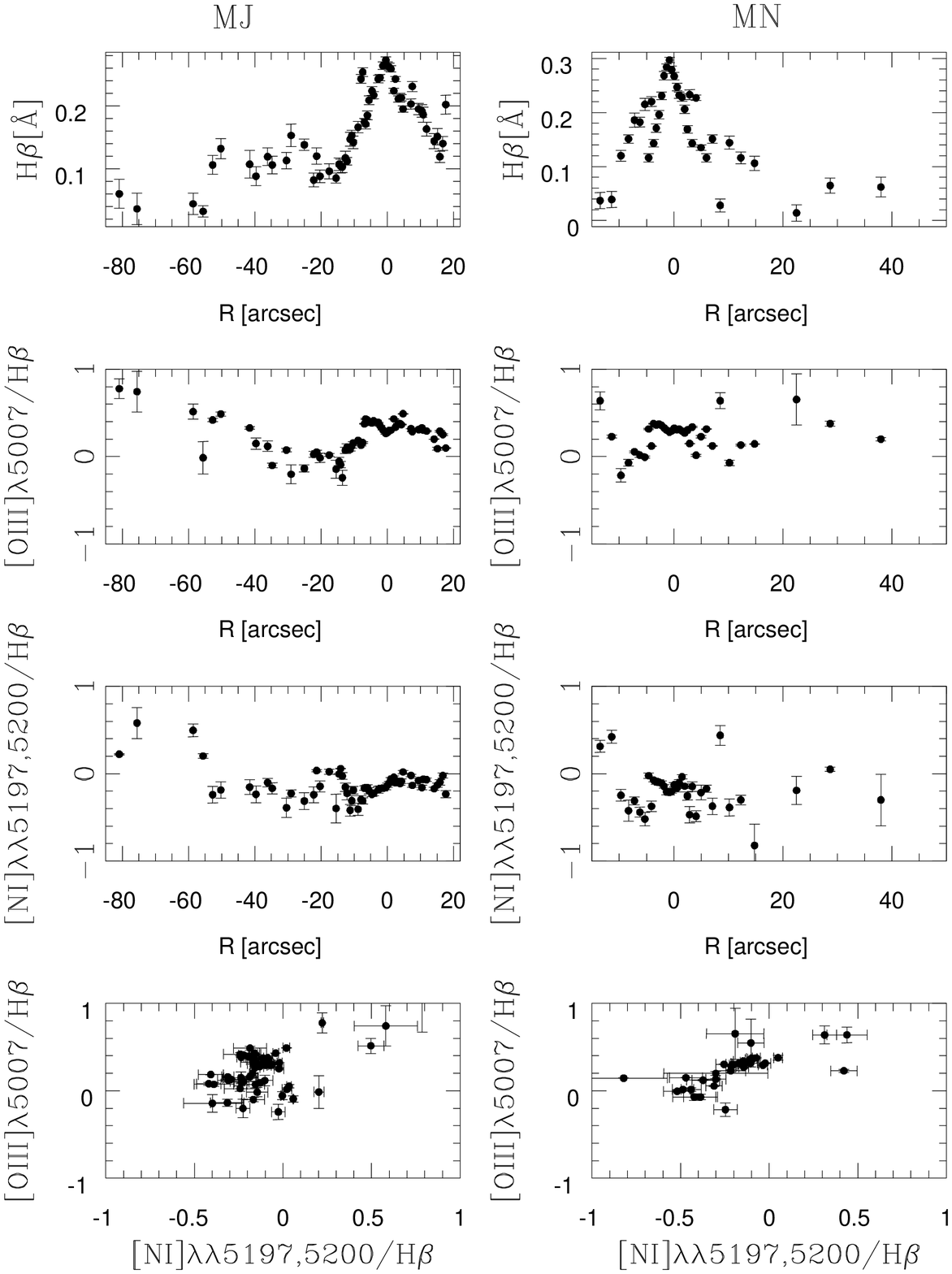}
\end{tabular}
\end{center}
\caption{The gas kinematics both along major and minor axes
  of \object{NGC4125} are shown on the left; open and filled symbols 
  show different sides of the galaxy. The emission lines (in 
logarithmic units) along the
  major and minor axes of \object{NGC4125} are show on the right. From
  the first row to the third row, we show the H$\beta$
  emission EW, the [OIII]$\lambda$5007/H$\beta$ and the
  [NI]$\lambda\lambda$5197,5200/H$\beta$ ratios (in logarithmic units) as a function of
distance from the center. In the bottom panels the diagnostic plots 
  [NI]$\lambda\lambda$5197,5200/H$\beta$ against
  [OIII]$\lambda$5007/H$\beta$ in logarithmic units are shown. \label{fig_gas}}
\end{figure*}

Compared to the kinematic profiles of the stars, the velocity curves
of the gas along the axes shows stronger gradients, see left panel of
Fig. \ref{fig_gas}, and presents a maximum of about 240km/s at 10
arcsec. Warm gas has been found already by
\citet{1983IAUS..100..311B}.  Our folded gas velocity curve is not
perfectly axisymmetric, but less distorted than what reported there.
Finally, the rotation curve implied by the stellar dynamical modeling
is much higher ($\approx 300$ km/s) than the measured
velocities. Therefore the gas is not following simple circular motions
and/or is not settled in a regular disk.  The gas velocity dispersion
is high in the inner 20 arcsec ($\approx 160$ km/s after correction
for instrumental broadening) and unconstrained at larger radii, where
it is less or comparable to our spectral resolution. Figure
\ref{fig_gas} shows also the equivalent widths (in {\AA}) of the
emission line H$\beta$ and the ratios
[NI]$\lambda\lambda$5197,5200/H$\beta$ and
[OIII]$\lambda$5007/H$\beta$ in logarithmic units.  The data along the
major and minor axes are shown separately.  The
[NI]$\lambda\lambda$5197,5200/H$\beta$ versus
[OIII]$\lambda$5007/H$\beta$ diagnostic diagram are shown at the
bottom of right panel.  As it can be seen from the plot, the ratios
[OIII]$\lambda$5007/H$\beta$ span the range from 0 to 0.5 and the
ratios [NI]$\lambda\lambda$5197,5200/H$\beta$ go from -0.5 to 0.0.
According to \citet{2009arXiv0912.0275S}, this is the region driven by
LINER-like emission.

\emph{\object{NGC7619}:} Our kinematic profiles for \object{NGC7619} extend to almost 1.6 a$_e$
along the major axis and 2 b$_e$ along the minor axis, further out than
previous works, for example 
\citet{1995ApJ...448..119F,1998A&AS..130..267L} extended only up to
2/3 R$_e$. Our data (circles in Fig.2) agree, within the error, with
the measurement of \citet{1995ApJ...448..119F} (squares).  From the
kinematic profiles, we see that the velocity dispersion of the galaxy
has a shallow gradient inside 10 arcsec and remains nearly flat towards
large radii. The rotational velocity also remains almost constant from 
8~arcsec out to 60~arcsec.  Furthermore, slow rotation along the minor axis
reveals that the galaxy is possibly triaxial \citep{1984AJ.....89..356B}.
Concerning the lines strength profiles, the galaxy shows constant
H$\beta$ index from the center to the outer regions and it has a steep
Mg\emph{b} gradient.

\section{Stellar populations}
\label{sec_stelpop}

\subsection{Model and method}
\label{sec_SPPmod}

In this section, we use the stellar population models of TMB03
to derive the age, total metallicity and element abundance gradients 
along the major and minor axes from the measured lines indices of 
our galaxies. The TMB03 models cover ages
between 1 and 15 Gyr, metallicities between 1/200 and 3.5 solar.
Furthermore, the models take into account the effects on the Lick
indices by the variation of element abundance, hence, give Lick
indices of simple stellar populations not only as the function of age
and metallicity, but also as the function of the $\alpha/Fe$ ratio.
The age, total metallicity and element abundance can be derived from a
comparison of selected line strength indices with SSP models TMB03.

The traditional and effective method of studying stellar population
properties uses diagrams of different pairs of Lick indices
\citep{2005ApJ...621..673T}.  The $H\beta$ versus [MgFe]\'{} pair
diagram is selected as the best age indicator because $H\beta$ is
sensitive to warm turnoff stars and [MgFe]\'{} index is considered as
the best detector of metallicity since it does not depend on abundance
ratio variations. Using $H\beta$ versus [MgFe]\'{} can break the
age-metallicity degeneracy.  The $<Fe>=(Fe5270+Fe5335)/2$ versus
Mg\emph{b} pair usually is considered as the best indicator of the
abundance of populations.  However, in this study, following
\citet{2009arXiv0910.5590S} we use a simple and perhaps more accurate
method- $\chi^{2}$ minimization:
\begin{equation}
\label{eq_chi}
\chi^{2}=\sum_{index[i]}\frac{(index_{i[ob]}-index_{i[mod]})^{2}}{(\sigma_{i[ob]})^{2}}
\end{equation}
where $index_{i[ob]}$ and $index_{i[mod]}$ represent the
$i^{th}$ observational indices and model indices respectively,
$\sigma_{i[ob]}$ is the observational uncertainty of $i^{th}$
indices. The best
fitting age, metallicity and $\alpha/Fe$ can be derived by
finding the minimum $\chi^{2}$ of all selected lines
indices to the SSP models. We chose H$\beta$, Mg\emph{b},
$Fe_{5015}$, $Fe_{5270}$, $Fe_{5335}$ and $Fe_{5406}$ as the
indicators. Further, in order
to improve the precision of the stellar properties using the
$\chi^{2}$ minimization method we interpolated the tabulated
indices of TMB03 on steps of 0.1 Gyr in age, 0.02 in
metallicity and 0.05 in $\alpha/Fe$.

\begin{figure*}
\begin{center}
\begin{tabular}{c}
\includegraphics[width=0.4\textwidth]{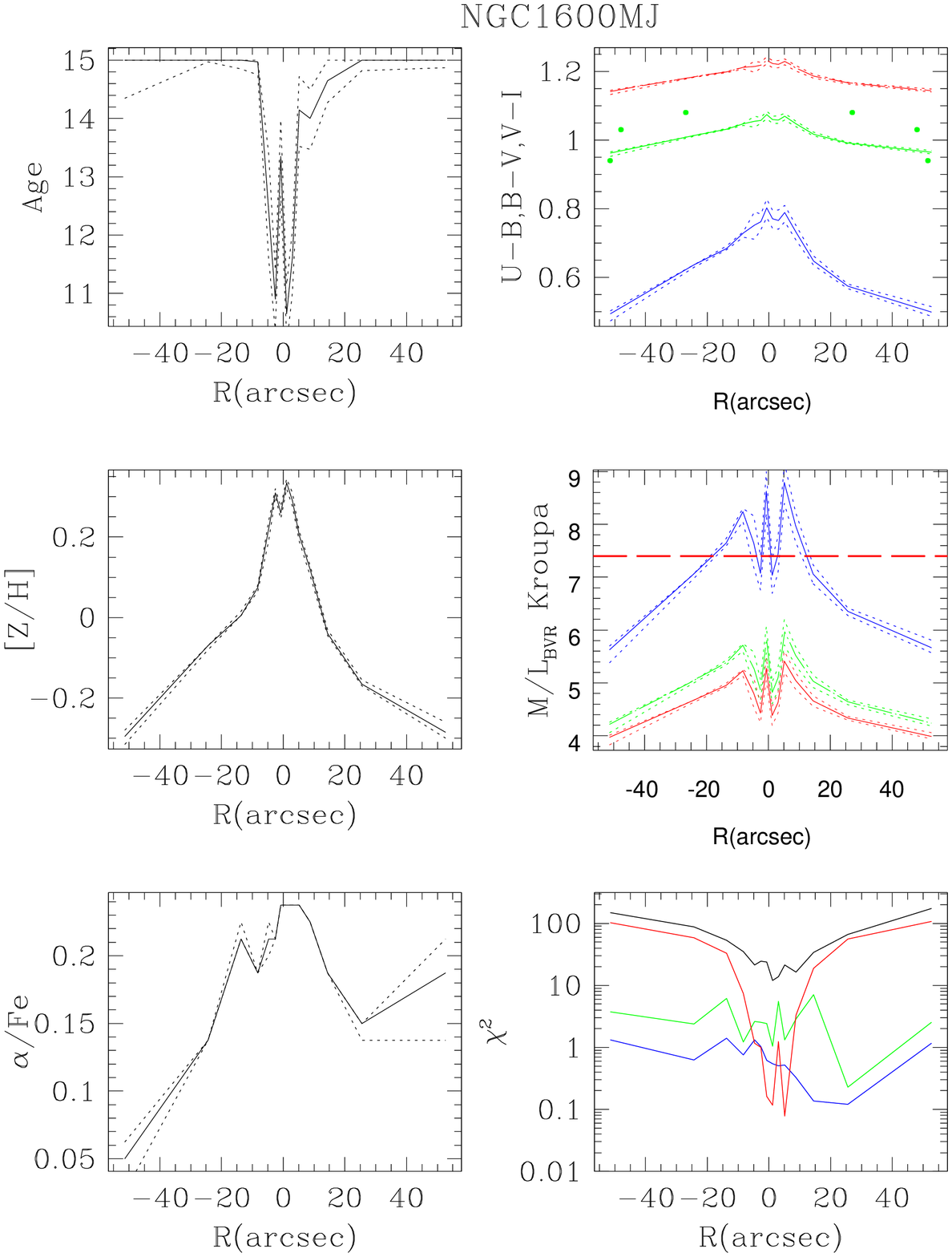}
\includegraphics[width=0.4\textwidth]{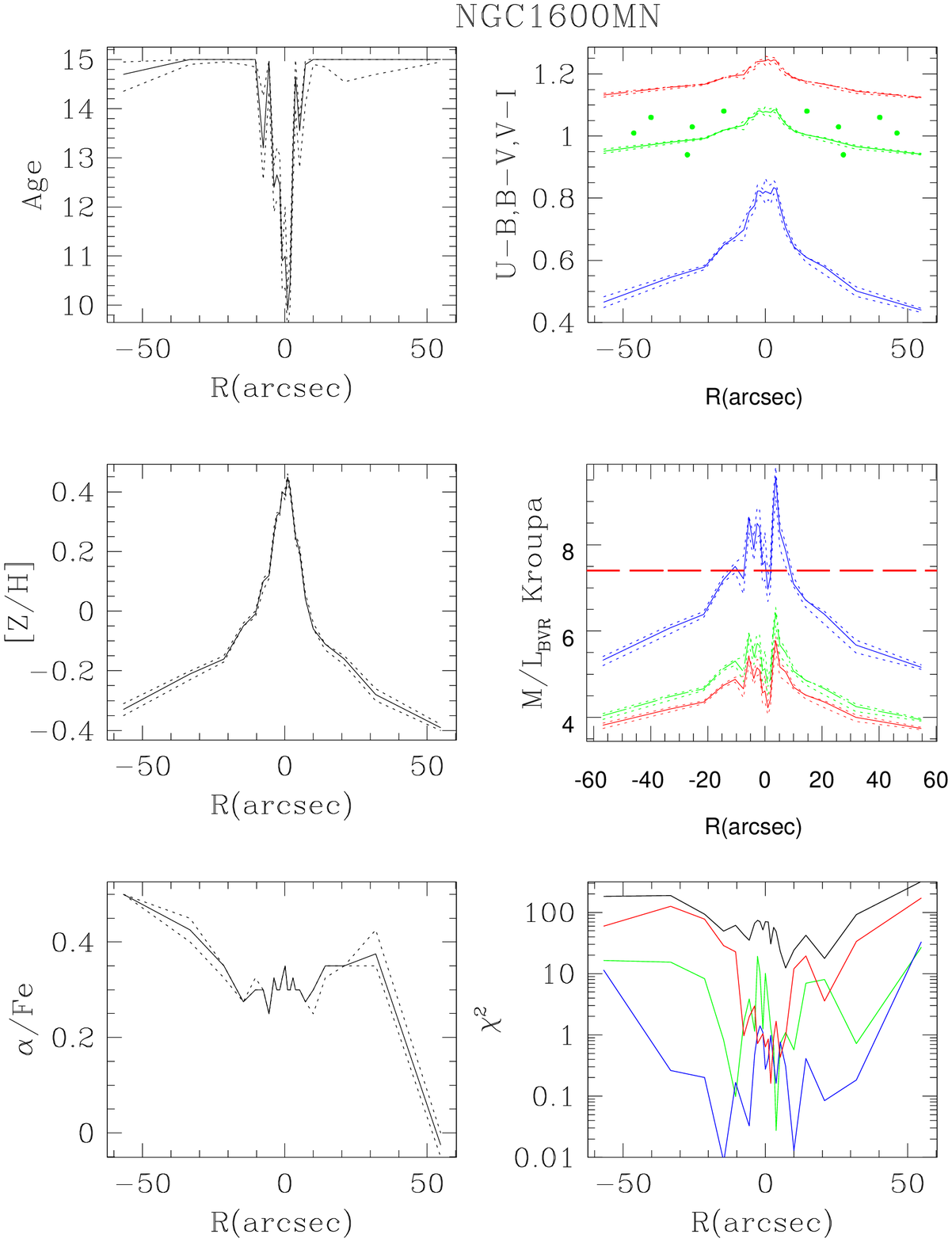}\\
\end{tabular}
 \end{center}
 \caption{The best fitting SSP equivalent age, metallicity and
   element abundance ratio are showed in the left in each
   plot. The right top of each plot shows the Johnson
   broad band U-B, B-V, V-I colors profiles, the blue and red
   solid lines stand for U-B and B-V color respectively;
   the black solid lines indicate the V-I color. For
   \object{NGC4125} and \object{NGC7619} the measured V-I color of
   \citet{1989A&A...217...35B} are shown as dots. In the case of
   \object{NGC1600}, the aperture B-V color taken from
   \citep{1973ApJ...183..711S} are indicated.  The M/L$_{BVI}$
   are shown in the mid of the right columns, the blue, green and red solid
   lines indicate the M/L in the B, V and I band respectively; the red dashed
   line shows the best-fit dynamical M/L in the I band for \object{NGC4125} and \object{NGC7619} 
   and in the R  band for \object{NGC1600}; the
   minimized $\chi^{2}$ of selected line strengths are
   presented in the bottom panel on the right, red, blue and green lines 
   present the minimized $\chi^{2}$ of H$\beta$, Mg\emph{b} and Fe$_{5015}$ 
respectively;
   while the black lines show the total minimized $\chi^{2}$. 
   \label{fig_SSP}}
 \end{figure*}

 \begin{figure*}
 \begin{center}
 \begin{tabular}{c}
 \includegraphics[width=0.4\textwidth]{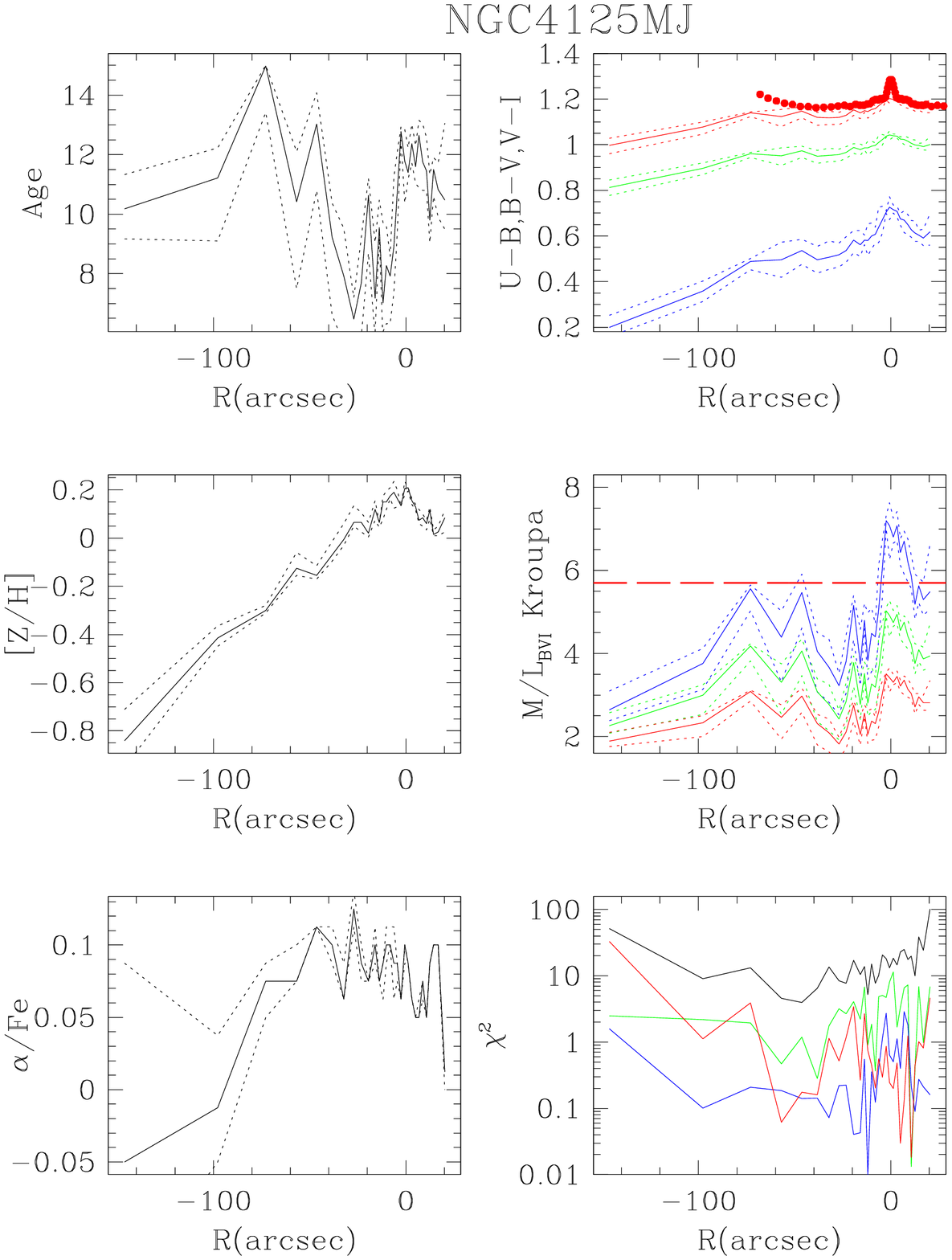}
 \includegraphics[width=0.4\textwidth]{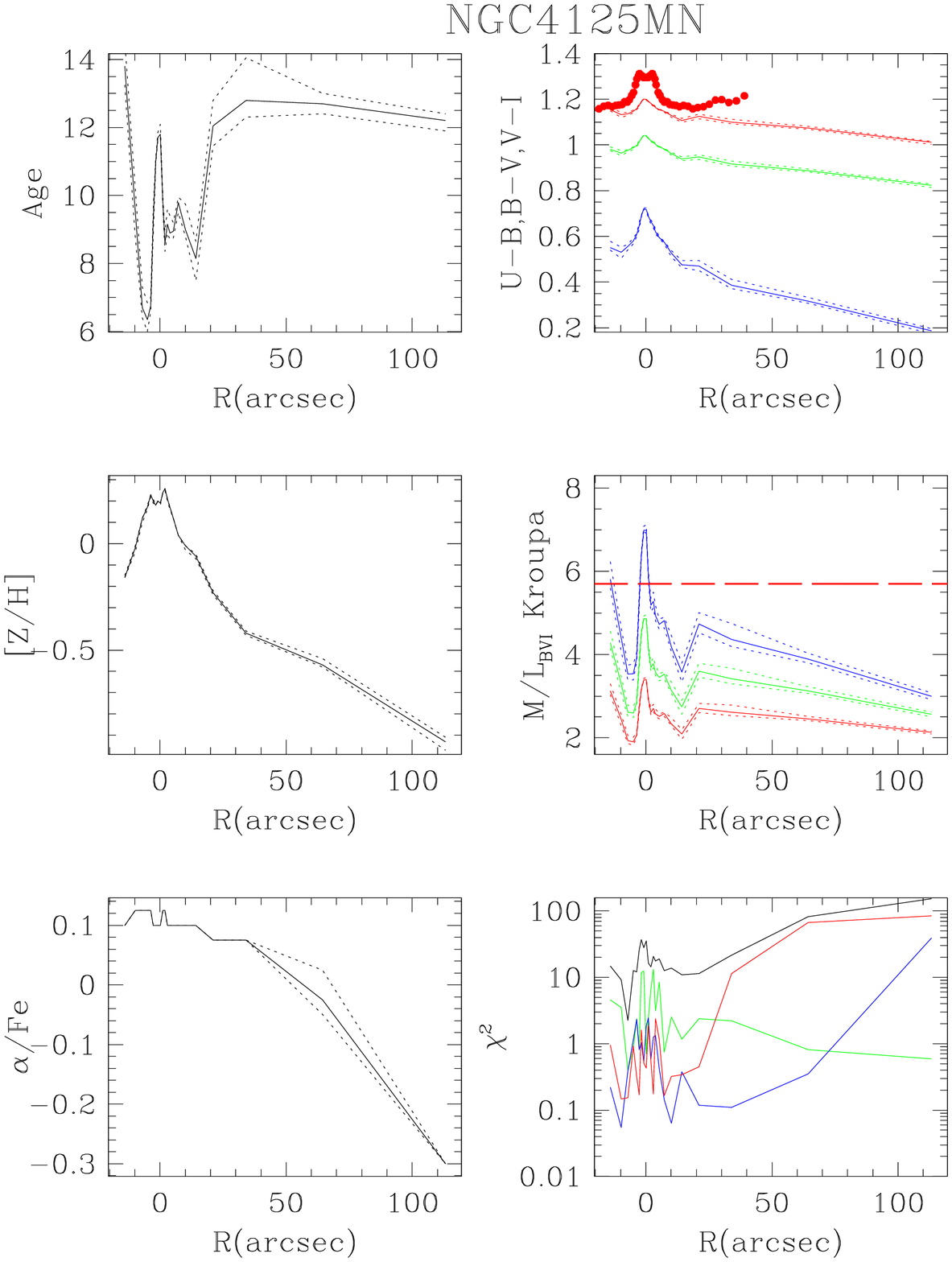}\\
 \end{tabular}
 \end{center}
 \end{figure*}

\addtocounter{figure}{-1}
\begin{figure*}
\begin{center}
\begin{tabular}{c}
\includegraphics[width=0.4\textwidth]{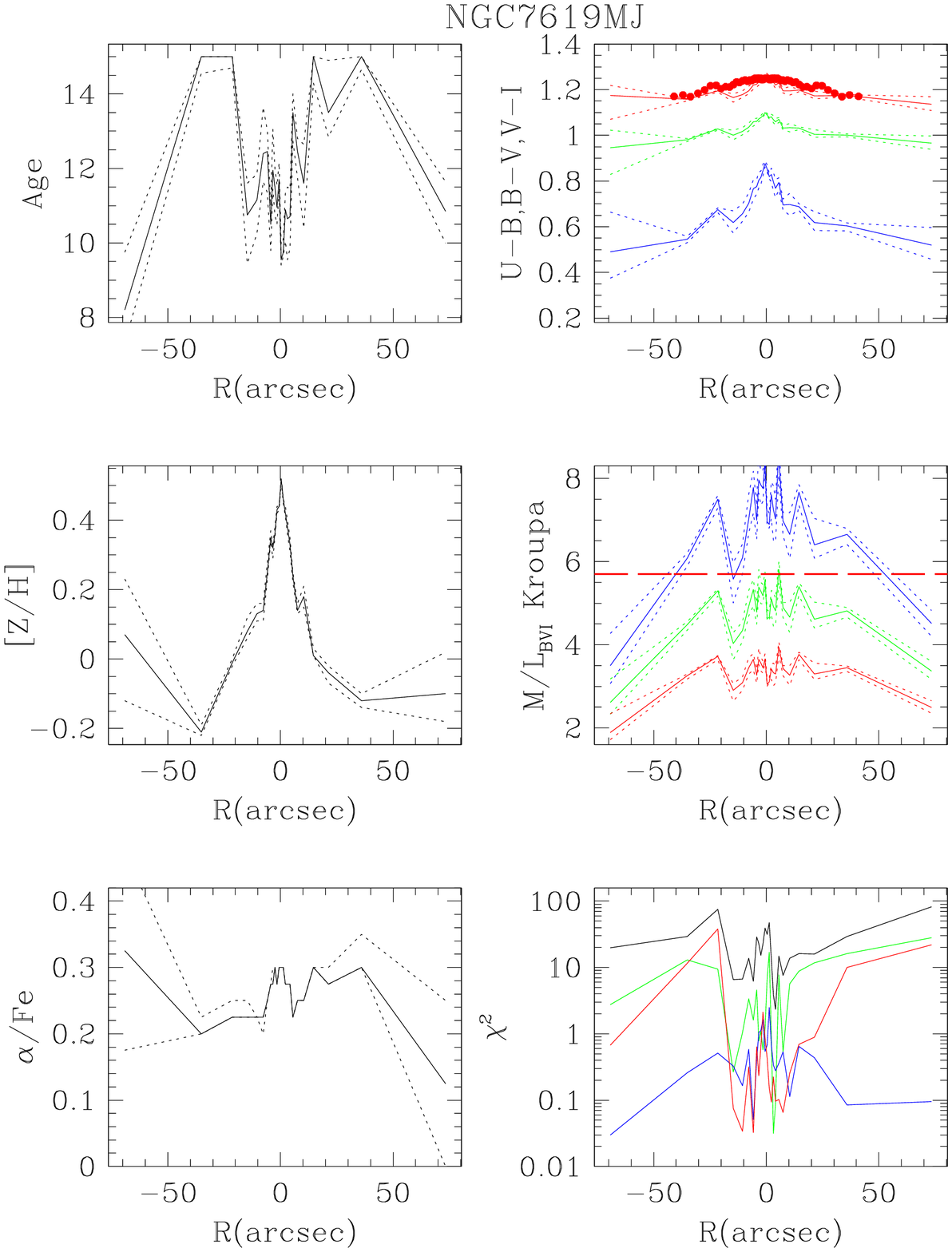}
\includegraphics[width=0.4\textwidth]{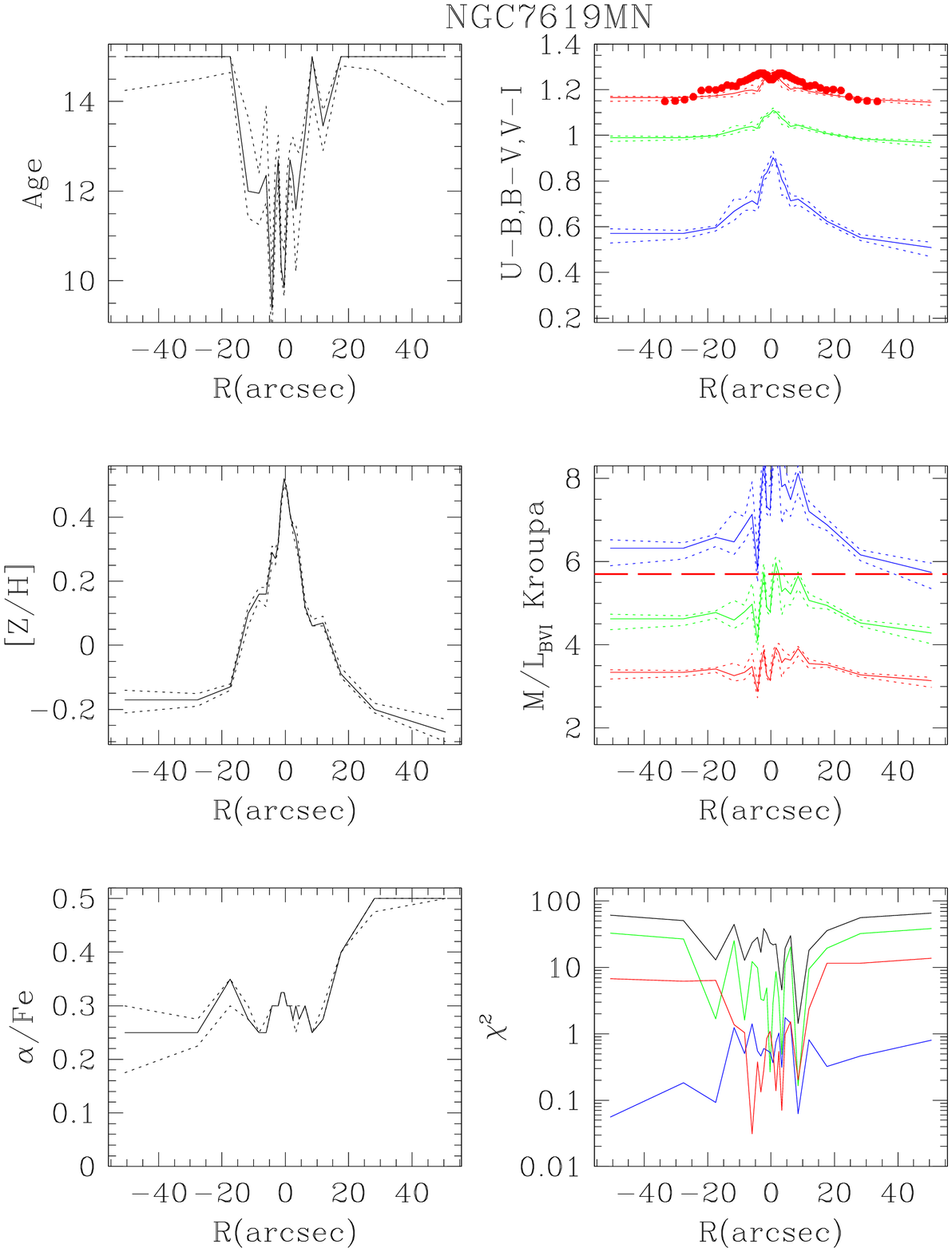}
\end{tabular}
\end{center}
\caption{Continued.}
\end{figure*}

\subsection{Ages, metallicities and $\alpha/Fe$ ratio profiles}
\label{sec_SPPresults}
 
The model predicted lines strength profiles are shown in Fig.
\ref{fig_lin} with solid lines. The derived quantities (age,
metallicity, element abundance, colors, mass-to-light ratios, and
resulting $\chi^2$) are shown in Fig. \ref{fig_SSP}. Here we average
together every four points to reduce the scatter. The theoretical
lines strength indices match well the measured parameters in the inner
regions of the galaxies, although an acceptable value of
$\chi^{2}\approx 3$ is obtained seldom (see Fig. \ref{fig_SSP}),
possibly indicating a global underestimation of our statistical
errors, or, more probably, the presence of systematic effects. In
particular, in the case of \object{NGC1600}, the large values of $\chi^{2}$ are
mainly driven by the significantly low H$\beta$ (Fig. \ref{fig_lin}
and red solid lines in the right bottom panel of Fig.
\ref{fig_SSP}). The large values of $\chi^{2}$ in the outer regions
along the minor axis of \object{NGC4125} are caused by the lower H$\beta$
strength.  In the case of \object{NGC7619}, the large values of $\chi^{2}$ in
the outer parts along both the major and the minor axes are mainly due
to large divergence of Fe5015 (green solid lines).  The SSP equivalent
age, metallicity and element abundance ratio along the major and minor
axis of our galaxies are also plotted on the left side of each panel.
In each panel the solid lines indicate the fitting parameter profiles
and the dotted lines show the 1 $\sigma$ errors. As it can be seen in
Fig.~\ref{fig_SSP}, all the galaxies are $[\alpha/Fe]$ overabundant,
at the value of 0.25, 0.1 and 0.3 dex for \object{NGC1600}, \object{NGC4125} and \object{NGC7619}
respectively. The $\alpha/Fe$ ratios are fairly constant in the
central regions, with hints for a decrease in the outer parts. Table
\ref{tab_grad} lists the slope of $\triangle \mbox{index}/\triangle
log(r)$ of our galaxies which were derived by fitting a straight line
to our measured data points.

 \begin{table*}
 \caption{Line strength and total metallicity gradients:$\triangle$ index/ $\triangle$ log (r). \label{tab_grad} }
 \begin{center}
 \begin{tabular}{lccccccccccccr}
 \hline
 Galaxy    & Range    &Mgb      & Fe5015  &Fe5270  &Fe5335 & $<Fe>$  & Fe5406   &H$\beta$   &Z/H   &$\alpha/Fe$ &Age\\
 Name     & ($R_e$)     &Slope    &Slope    &Slope   &Slope  &Slope    &Slope     &Slope      &Slope &Slope      &Slope\\
          &          &$\pm $   &$\pm$    &$\pm $  &$\pm$  &$\pm $   &$\pm$     &$\pm$      &$\pm$ &$\pm$       &$\pm$\\
 \hline
 \object{NGC1600}& 0-0.5     &-0.787    &-0.339   &-0.179 &-0.196 &-0.187 &-0.207 &-0.125  &-0.342  &-0.042  &1.32 \\
        &           &0.067     &0.082    &0.036  &0.053  &0.038  &0.094  &0.028   &0.029   &0.012   &0.43\\
        & 0-1       &-0.742    &-0.537   &-0.268 &-0.222 &-0.245 &-0.236 &-0.172  &-0.288  &-0.044  &1.398\\
        &           &0.069     &0.085    &0.037  &0.046  &0.036  & 0.029 & 0.033  & 0.028  &0.009   &0.343\\
 \object{NGC4125} &0-0.5      &-0.378    &-0.328   &-0.215 &-0.267 &-0.241 &-0.121 &0.224   &-0.225  &-0.037  &-3.527\\
        &           &0.021     &0.036    &0.038  &0.020  &0.024  &0.017  &0.020   &0.013   &0.006   &1.234\\
        &0-1        &-0.464    &-0.405   &-0.260 &-0.319 &-0.289 &-0.185 &0.173   &-0.235  &0.008   &-2.464\\
        &           &0.023     &0.035    &0.033  &0.020  &0.021  &0.019  &0.020   & 0.014  &0.006   &1.250\\
 \object{NGC7619} &0-0.5      &-0.711    &-0.571   &-0.192 &-0.352 &-0.272 &-0.299 &0.078   &-0.343  &0.041   &-1.420\\
        &           &0.049     &0.082    &0.035  &0.038  &0.028  &0.036  &0.021   &0.028   &0.010   &0.614\\
        &0-1        &-0.787    &-0.953   &-0.280 &-0.437 &-0.359 &-0.264 &-0.026  &-0.327  &-0.034  &2.143\\
        &           &0.048     &0.103    &0.035  &0.037  &0.028  &0.036  &0.039   &0.024   &0.011   &0.497\\
\hline
\end{tabular}
\end{center}
\end{table*}

 The nuclear (inside R$_e$/8) parameters were derived by
 \citet{2000AJ....119.1645T} and \citet{2005MNRAS.358..813D}. 
 In order to compare 
 to values available in the literature we calculate the mean
 parameters inside R$_e$/8.  For \object{NGC1600} and \object{NGC7619}
 \citet{2000AJ....119.1645T} derived the nuclear parameters [age,
 Z/H, $\alpha/Fe$] with [8.6 $\pm$ 1.7 Gyr, 0.35 $\pm$ 0.05, 0.22 $\pm$
 0.02] and [14.8 $\pm $2.3 Gyr, 0.2 $\pm$ 0.03, 0.18 $\pm$ 0.01]
 which compare to our value of 
[ 13.1 Gyr $\pm$ 1.1, 0.35 $\pm$ 0.03, 0.317 $\pm$ 0.001] and 
[ 11.7 Gyr $\pm$ 0.9, 0.25 $\pm$ 0.02, 0.284 $\pm$ 0.001] 
respectively. In the case of \object{NGC4125},
 \citet{2005MNRAS.358..813D} obtained [age, Z/H, $\alpha/Fe$] with [5.9
 $\pm $3 Gyr, 0.32 $\pm$ 0.1, 0.1] while we find 
[ 10.1 Gyr $\pm$ 0.6, 0.12 $\pm$ 0.02, 0.095 $\pm$ 0.001].

 \subsection{Color and M/L profiles}
\label{sec_colors}

The Johnson broad band U-B, U-V, B-V, V-R, V-I, V-K, J-K, J-H, H-K
color and M/L ratio in B, V, R, I, J, H, and K bands profiles were
calculated using the Kroupa initial mass function
\citet{1995ApJ...453..350K} with the help of the SSP models
\citep{1998MNRAS.300..872M}. For a clear presentation in the figures,
we only show the U-B, B-V and V-I color and the mass to light ratios
in B, V, I band in this papers.  On the top right panels in Fig.
\ref{fig_SSP}, the blue and red solid lines stand for U-B and B-V
color respectively; the black lines indicate the V-I color.  For
\object{NGC4125} and \object{NGC7619}, the measured V-I color are also over plotted with
solid dots.  For \object{NGC1600}, unfortunately, we do not have reliable broad
band color profiles and we take the aperture B-V color from
\citet{1973ApJ...183..711S}.  As it can be seen from the plots, the
models predicted color profiles agree reasonably well with the
measured colors except for small discrepancies. We note that
\citet{2009MNRAS.394L.107M} discuss the limitations of current SSP
models in predicting accurate colors. In particular, the B-V colors of
old ($\sim 12$ Gyr), solar metallicity SSP models are expected to be
-0.09 mag bluer than what used here, while the differences are smaller
for the redder colors. In addition, if our estimated ages are too
small because of the presence of a second younger component
\citep{2007MNRAS.374..769S}, then the predicted SSP colors would be biased and bluer than they should. Moreover, SSP models do not account for the presence of
dust, which could be present especially at the centers of our
galaxies. The middle right of Fig.7 shows the theoretical M/L ratio in
B, V, I bands; the blue and green lines display the M/L ratios in B and
V bands respectively; while the red lines present the I band M/L
ratios. The red long dashed lines show the dynamical 
M/L in the I band for \object{NGC4125} and \object{NGC7619}, and in the R band for \object{NGC1600}
(see Sect. \ref{sec_dynamics}). 
The minimized $\chi^{2}$ of selected lines strength are
presented in the bottom panel on right, red, blue and green lines
present the minimize $\chi^{2}$ of H$\beta$, Mg\emph{b} and
Fe$_{5015}$ respectively; while the black lines show the total
minimized $\chi^{2}$. All of the our giant elliptical galaxies show a
red sharp peak, mainly due to metallicity gradients
\citep{1989PhDT.......149P}.

\section{Line strengths and the local escape velocity}
\label{sec_dynamics}

As discussed in the Introduction, we want to exploit the link
that the galaxy formation process has established between the Mg\emph{b} line
strength and the local escape velocity to constrain the density
profile of dark matter halos in the outer regions of galaxies. 
The $V_{esc}$ is given by:\\
\begin{equation}
\label{eq_vesc}
V_{esc} = \sqrt{2|\Phi|},
\end{equation}
and therefore at each radius it is sensitive to the total mass density
profile up to very large radii; for example, in the case of a
spherical density distribution $\rho$ the gravitational potential $\Phi$ is:
\begin{equation}
\label{eq_phispher}
\Phi(r)=-G\left[ \frac{M(<r)}{r}+ 4\pi \, \int_r^\infty \rho(r')r'dr'\right],
\end{equation}
where $M(<r)=4\pi\int_0^r r'^2\rho(r')dr'$. In the following we will 
correlate Mg\emph{b} measured at the projected distance $r$ with 
$V_{esc}$ computed at the intrinsic distance $r$. As noted by  
\citet{2009MNRAS.398.1835S}, we see no difference if instead we compute the 
projected quantity:
\begin{equation}
V_{esc,pr} = \frac{\int V_{esc} \, \rho_\ast \mathrm{d}z}{\int \rho_\ast \,
\mathrm{d}z}
\label{eq_vescproj}
\end{equation}
where $z$ is the line-of-sight. Moreover, having in mind the physical
scenario described in the Introduction, where more enrichment in
Mg\emph{b} is achieved if more gas is retained, one would expect the
difference $\Delta V$ between $V_{esc}(r) $ and the average velocity
of stars and gas at that location $r$ to play the crucial role. In
fact, $\Delta V$ happens to be proportional to $V_{esc}$, because
elliptical galaxies are virialized systems.

\subsection{Dynamical models}
\label{sec_dynmet}

We use the axisymmetric Schwarzschild's orbits superposition technique 
\citep{1979ApJ...232..236S} to derive the gravitational potential
profiles. Thereby, also the stellar
mass-to-light ratio, the internal orbital structure as well as
the velocity anisotropy of the galaxies can be determined.
Here, we only briefly present the basic steps,
details about our implementation of the
method are given in \citet{2004MNRAS.353..391T,2005MNRAS.360.1355T}.\\
(1): We first determine the luminosity density from the surface
brightness profile.  The photometric data for the modelling are
deprojected into a three-dimensional, axisymmetric distribution 
with specified
inclination using the program of \citet{1999MNRAS.302..530M}.\\
(2): The total mass distribution of the galaxies consists of the stellar mass
density and a dark matter halo:
\begin{equation}
\label{eq_rho}
\rho = {\Upsilon}{\nu}+ \rho_{DM},
\end{equation}
where $\Upsilon$ is the stellar mass-to-light ratio, and ${\nu}$ is the
deprojected stellar luminosity density.\\
(3): The gravitational
potential $\Phi$ can be derived by integrating Poisson's equation once
the total mass profile is obtained. Thousands of
orbits are calculated in this fixed potential.\\
(4): The orbits are superposed to fit the observed LOSVDs,
following the luminosity density constraint. The maximum entropy
technique of \citet{1988ApJ...327...82R} is used to fit the kinematics
data by maximizing the function:
\begin{equation}
\label{eq_entropy}
$\^{S}$ = S - {\alpha}\chi_{kin}^{2},
\end{equation}
where S is an approximation to the Boltzmann entropy and
${\chi}^{2}$ is the sum of the squared residuals to the kinematic
data. The smoothing parameter ${\alpha}$ controls the
influence of the entropy S on the orbital weights, see
\citet{2004MNRAS.353..391T} for more details. 

Two different types of dark matter halo distribution were tried to
recover the mass profiles.  First, the NFW halo profile
\citep{1996ApJ...462..563N}:
\begin{equation}
\label{eq_NFW}
\rho_{NFW}(r)=\frac{\rho_s}{(r/r_s)(1+r/r_s)^{2}},
\end{equation}
where $\rho_s$ is the characteristic density of the halo and
$r_s$ is a characteristic radius. One further defines the so-called 
concentration parameter $c$ of the halo that
is related to $r_s$ and the viral radius $r_{200}$ via $r_s = r_{200}/c$.
The potential generated by this density distribution is given by:
\begin{equation}
\label{eq_phi}
\Phi(r) = 4{\pi}G\rho_s{r_s}^{2}\frac{r_s}{r}\ln(r+\frac{r}{r_s}),
\end{equation}
where G is the gravitational constant. The second halo used is the
cored logarithmic (LOG) halo profile \citep{1987gady.book.....B} , 
 that reads as:
\begin{equation}
\label{eq_log}
\rho_{LOG}(r) = \frac{
{V_c}^{2}}{4{\pi}G}\frac{3{r_c}^{2}+r^{2}}{({r_c}^{2}+r^{2})^{2}}.
\end{equation}
This gives a roughly constant circular velocity $V_c$ and a
flat density core inside $r < r_c$. The potential is given by:
\begin{equation}
\label{eq_philog}
\Phi(r) = \frac{V_C^2}{2} \ln(r^2 + r_c^2)
\end{equation}
Both the NFW and the LOG potential give divergent mass profiles when
integrated to infinity. In our numerical implementation the potential
is computed up to 10 effective radii and extrapolated Keplerian at
larger radii.  In order to investigate further the physical extent of
such halos we introduce a cut-off radius $r_{cut}$ and modify the
density profile given by Eq. \ref{eq_log} as follows:
\begin{equation}
\label{eq_rhocut}
\rho_{cut}(r)=\rho_{LOG}(r)\times \exp \left( -\frac{r^2}{r_{cut}^2} \right).
\end{equation}

\subsection{Model results for \object{NGC1600}, \object{NGC4125} and \object{NGC7619}}
\label{sec_dynres}

\emph{Photometric data}: We produced isophotal fits separately on 
the HST and the ground based images using the code of \cite{BM87}.
For \object{NGC4125} and \object{NGC7619}, the surface
brightness profiles consist of HST F814W(I) filter images and SDSS $i$
band images, scaled to I band.  The photometric data of \object{NGC4125}
extended to 216~arcsec, about 3.5 times of effective radius.  For
\object{NGC7619}, the photometric data extended to 167~arcsec, more than 5
times of its effective radius. In the case of \object{NGC1600}, the photometric
data came from HST images observed with the F555W(R) filter and scaled
to the profile of \citet{1990AJ....100.1091P}. They extend to 
200~arcsec, nearly 4.5 times of the effective radius. The photometric
data are deprojected into the 3d luminosity distribution 
using the program of \citet{1999MNRAS.302..530M}.
\emph{Kinematic data set}: The kinematics data both along the major axes
and minor axes derived in Sect. \ref{sec_kin} are used.

We tested models at three different inclinations (i = 70, 80, 90) 
for our galaxies, the best-fit models for all three galaxies are edge-on ($i=90$).  The dark halo
parameters for the best-fit logarithmic halos are given in 
Tab.~\ref{tab_dynres}. They fit very well to the dark matter determinations of
Coma early-type galaxies by \citet{Tho09}. 
We also tested dark matter halos following the NFW halo profile,
as in \citet{2005MNRAS.360.1355T}.
The best-fit NFW-halos for \object{NGC1600}, \object{NGC4125} and \object{NGC7619} 
require concentration parameters c = 5, 5.5, 7.5
respectively. The mass-to-light ratios derived with NFW profile are
similar to the mass-to-light ratios with logarithmic halo profile.
Fig. \ref{fig_SSP} shows that the mass-to-light ratios derived from
the SSP models with a Kroupa IMF  
are on average 45\% lower than the dynamical ones, similarly to
\citet{2006MNRAS.366.1126C}. A better agreement is achieved if a
Salpeter IMF is considered. The comparison with the mass profiles
derived from X-ray measurements for \object{NGC4125} and \object{NGC7619}
\citep{2006ApJ...636..698F} is good: our total masses within the last
kinematic point are just 10\% larger than the X-ray ones. For \object{NGC1600}
the discrepancy is larger, with X-ray masses underestimating the
stellar kinematics ones by 30 (NFW potential) or 40 \% (LOG
potential).

As a further step, we compute a number of models with varying
$r_{cut}$ from 5 to 10 times effective radii for each galaxy.  The
different cut-off radii of the dark matter halos do not have
significant effects on the dynamical parameters ($\Upsilon$, $V_c$,
$r_c$), nor the quality of the kinematic fit (see upper panel of
Fig. \ref{fig_chi}, where the self-consistent model without dark
matter is plotted at $r_{cut}=0$ and the model with no cutoff at
$r_{cut}=80$ kpc), as soon as models with $r_{cut}\ge 40$ kpc are
considered. However, they do have an effect, when we also take into
account the correlation $Mg_b-V_{esc}$.


\begin{table}
\caption{Halo circular velocity $V_c$, halo core radius $r_c$, average dark matter density
$<\rho_\mathrm{DM}>$ (within $2 \, r_\mathrm{eff}$) and stellar mass-to-light ratio
$\Upsilon$ (last column: photometric band) for the best-fit dynamical models of
\object{NGC1600}, \object{NGC4125} and \object{NGC7619} (no cutoff radius). \label{tab_dynres}}
\begin{center}
\begin{tabular}{lccccc}
\hline
Name   & $V_c$  & $r_c$ & $\log_{10} <\rho_\mathrm{DM}>$ & $\Upsilon$ & band\\
       & (km/s) & (kpc) & ($M_\odot/pc^3$) & ($M_\odot/L_{\odot}$) &\\
\hline
\object{NGC1600} &  $350^{+250}_{-50}$  & $31^{+62}_{-12}$ & $-2.1 \pm 0.5$ & $7.3 \pm 0.5$ & R\\
\object{NGC4125}  & $350^{+100}_{-100}$  & $32^{+7}_{-7}$ & $-2.2 \pm 0.1$ & $5.7 \pm 0.7$ & I\\
\object{NGC7619}  & $600^{+200}_{-150}$  & $49^{+9}_{-16}$ & $-2.0 \pm 0.2$ & $6.5 \pm 0.5$ & I\\
\hline
\end{tabular}
\end{center}
\end{table}

\subsection{Mgb, the local escape velocity and the size of dark matter halos}
\label{sec_vesc}

Next we explore the correlation between Mgb and the local escape
velocity by fitting log Mg\emph{b} and log $V_{esc}$ to a
straight-line, weighted by the errors. We computed errors on $V_{esc}$
by considering at every point the range of escape velocities allowed
by all models giving $\chi^2_{kin}\le \chi^2_{kin,min}+1$.  According
to eq.~(\ref{eq_phispher}) the outer halo-profile has a considerable
effect on $V_{esc}$ and, hence, on the tightness of the correlation
between log Mg\emph{b} and log $V_{esc}$. Here, we explore shallower
halo profiles than probed by \citet{2009MNRAS.398.1835S} and use the
correlation between Mg\emph{b} and $V_{esc}$ to constrain the size of
dark matter halos.  We define a new parameter ${{\chi}_{tot}}^{2}$,
given by:
\begin{equation}
\label{eq_chitot}
\chi_{tot}^{2} = \chi_{Mgb}^{2}+\chi_{kin}^{2},
\end{equation}
where, $\chi_{Mgb}^{2}$ quantifies the goodness of fitting log
Mg\emph{b} and log $V_{esc}$ to a straight-line. The $\chi_{kin}^{2}$
is mentioned above, it quantifies the deviation between model and
observed kinematics.

We probed two different radial scalings: (1) correlating the measured
Mg\emph{b}(r) at radius $r$ with the escape velocity $V_{esc}(r)$ at
the same radius $r$ and (2) correlating Mg\emph{b} with the escape
velocity on the corresponding isophote (i.e. we compare the Mg\emph{b}
measured at a distance $b$ from the galaxy centre along the minor-axis
with $V_{esc}(b/\sqrt{q})$ in the equatorial plane and the Mg\emph{b}
measured at a distance $a$ from the galaxy centre along the major-axis
with $V_{esc}(a \sqrt{q})$, where $q=1-\epsilon$ is the apparent
flattening of the light-distribution). Scaling (2) gives a smaller
$\chi_{tot}^2$ and is discussed in the following. Using the scaling
(1), however, gives similar results with respect to the halo core
radii.

\begin{figure}
\begin{center}
\begin{tabular}{c}
\includegraphics[width=0.48\textwidth]{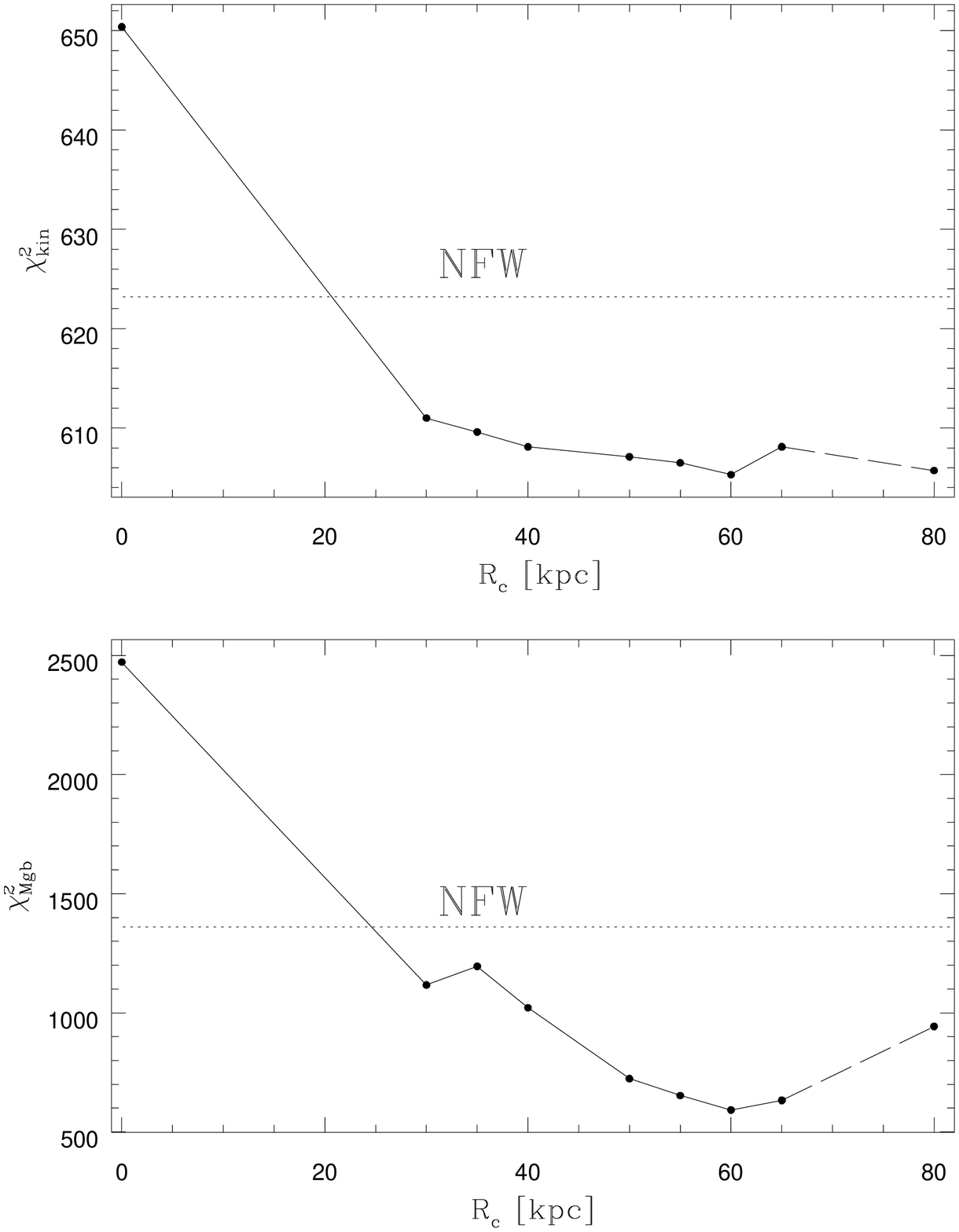}\\
\end{tabular}
\end{center}
\caption{The $\chi_{kin}^{2}$ as a function of cut-off radius 
summed over all three galaxies is shown in the 
upper panel and the $\chi_{Mgb}^{2}
$ is presented in bottom panel. The dotted lines shows the $\chi^{2}$ for the 
NFW halo profile, and the last point is the $\chi^{2}$ for the LOG dark halo 
profile without cutoff. \label{fig_chi} }
\end{figure}

The lower panel in Fig. \ref{fig_chi} shows $\chi_{tot}^{2}$ as a
function of the cut-off radius in kpc. The minimum scatter is achieved
for $r_{cut}=60$ kpc. This is 4.3 times the effective radii for
\object{NGC1600}, 9.5 times for \object{NGC4125} and 8.5 times for \object{NGC7619}. Lower cutoff
radii give slightly worse $Mgb-V_{esc}$ correlations, but the trend is
significant only in combination with $\chi^2_{kin}$. Cutoff radii
larger than 70 kpc have significantly higher $\chi^2_{tot}$. As
discussed above, this is driven only by the $Mgb-V_{esc}$ correlation.

\begin{figure}
\begin{center}
\begin{tabular}{c}
\includegraphics[width=0.48\textwidth]{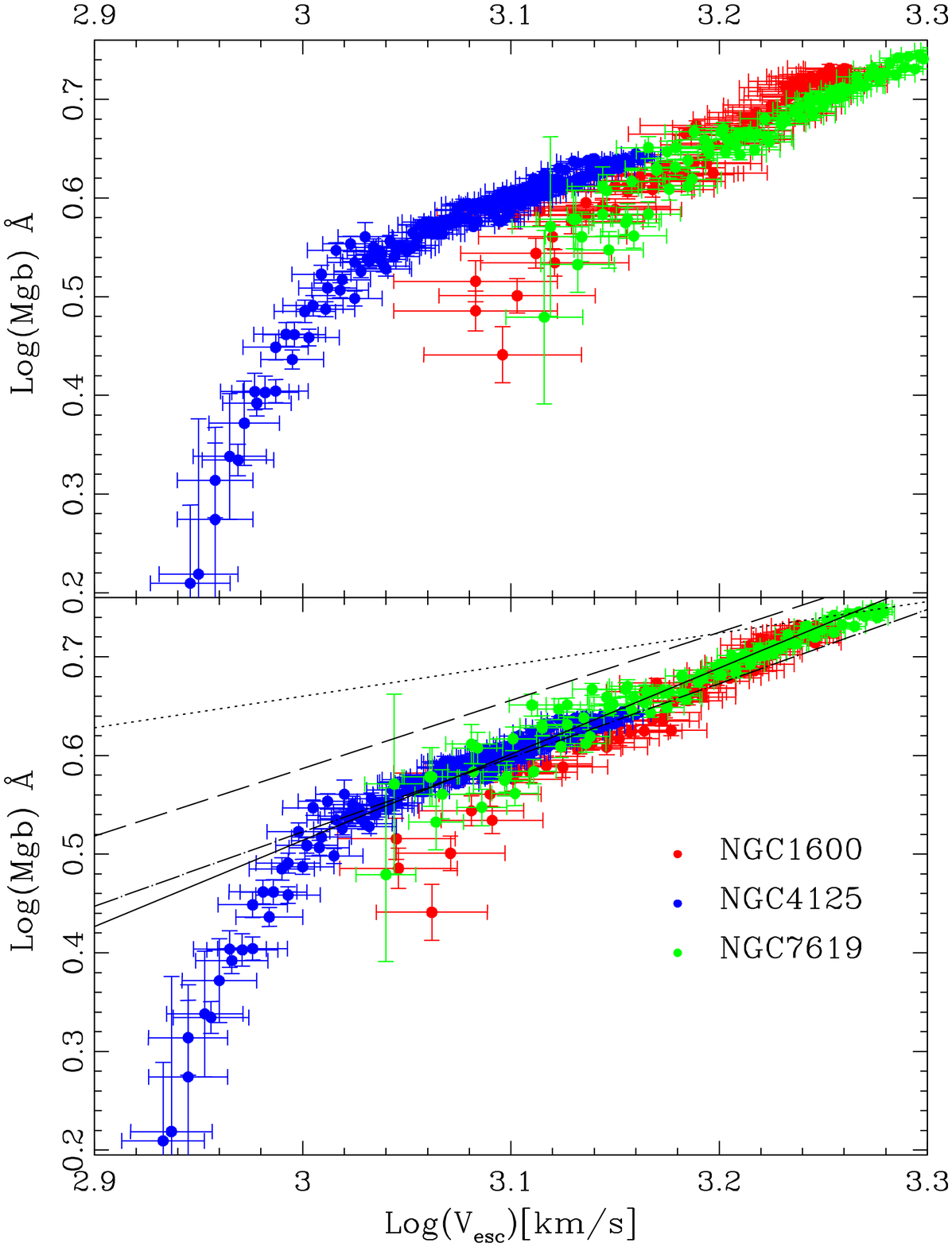}\\
\end{tabular}
\end{center}
\caption{The upper panel and bottom panel show the correlation between 
log(Mgb) and log($V_{esc}$) derived using LOG dark matter halos without 
cut off and cut off at 60 kpc respectively. Different colour refer to 
different galaxies. The full line is the corresponding best fit. The
dashed-dotted lines is computed for NFW dark matter halos.
The long-dashed line shows the regression obtained when models without 
dark matter halo are considered. The dotted line is taken from
\cite{2009MNRAS.398.1835S} without dark matter halos. 
\label{fig_vesc}}
\end{figure}

The \emph{Mgb-$V_{esc}$} correlation with the lowest $\chi^2_{tot}$ is
\begin{equation}
\log \mathrm{Mg}b = (0.873 \pm 0.008) \times \log \, V_{esc} - (2.105 \pm 0.024)
\end{equation}
 (shown in the bottom part of 
Fig. \ref{fig_vesc} by the solid line), the corresponding
\emph{$\chi_{Mgb}^{2}$} is 592 (for 368 data points).  The correlation
for the best-fitting NFW halo reads 
\begin{equation}
\log \mathrm{Mg}b = (0.629 \pm 0.004) \times \log \, V_{esc} - (1.314 \pm 0.013)
\end{equation}
 (shown by the long dashed-dotted
line), the corresponding \emph{$\chi_{Mgb}^{2}$} is 1360. The
regression obtained for models without dark matter is shifted to lower
$V_{esc}$, but is steeper than reported by
\cite{2009MNRAS.398.1835S}. For comparison, in the upper plot of Fig.  
\ref{fig_vesc} we show the \emph{Mgb-$V_{esc}$} relation obtained when 
the best-fitting logarithmic halo models without cutoff are used. 
 Note that we get the same cut-off radii if we use 
$[Z/H]-V_{esc}$ instead of \emph{Mgb-$V_{esc}$}. 

Figure \ref{fig_vesc} indicates that the simple linear relation
between $\log V_{esc}$ and $\log Mgb$ breaks down in the outer regions
of the galaxies investigated here. There, the measured value of $Mgb$
is much smaller than the linear relation determined in the inner parts
of the galaxies would predict. Due to the relatively low number of
points involved, the bending of the relation does not influence the
result of Fig. \ref{fig_chi}: the same optimal cut-off radius is found
if we do not consider points with $\log Mgb<0.5$. However, this might
be telling something about the formation process of ellipticals. It is
reminiscent of several recent observational  and 
theoretical findings \citep[][and references therein]{vanDokkum2010}
that suggest that the outer stellar envelopes of bright ellipticals
might be the result of later accretion of smaller objects on older
central cores formed in violent collapsing processes at high
redshift. In this case, the metallicity measured in the outer parts
would be set be the (lower) escape velocity of the accreted objects,
and not by the deeper potential well of the final giant elliptical.

\begin{figure}
\begin{center}
\begin{tabular}{c}
\includegraphics[width=0.48\textwidth]{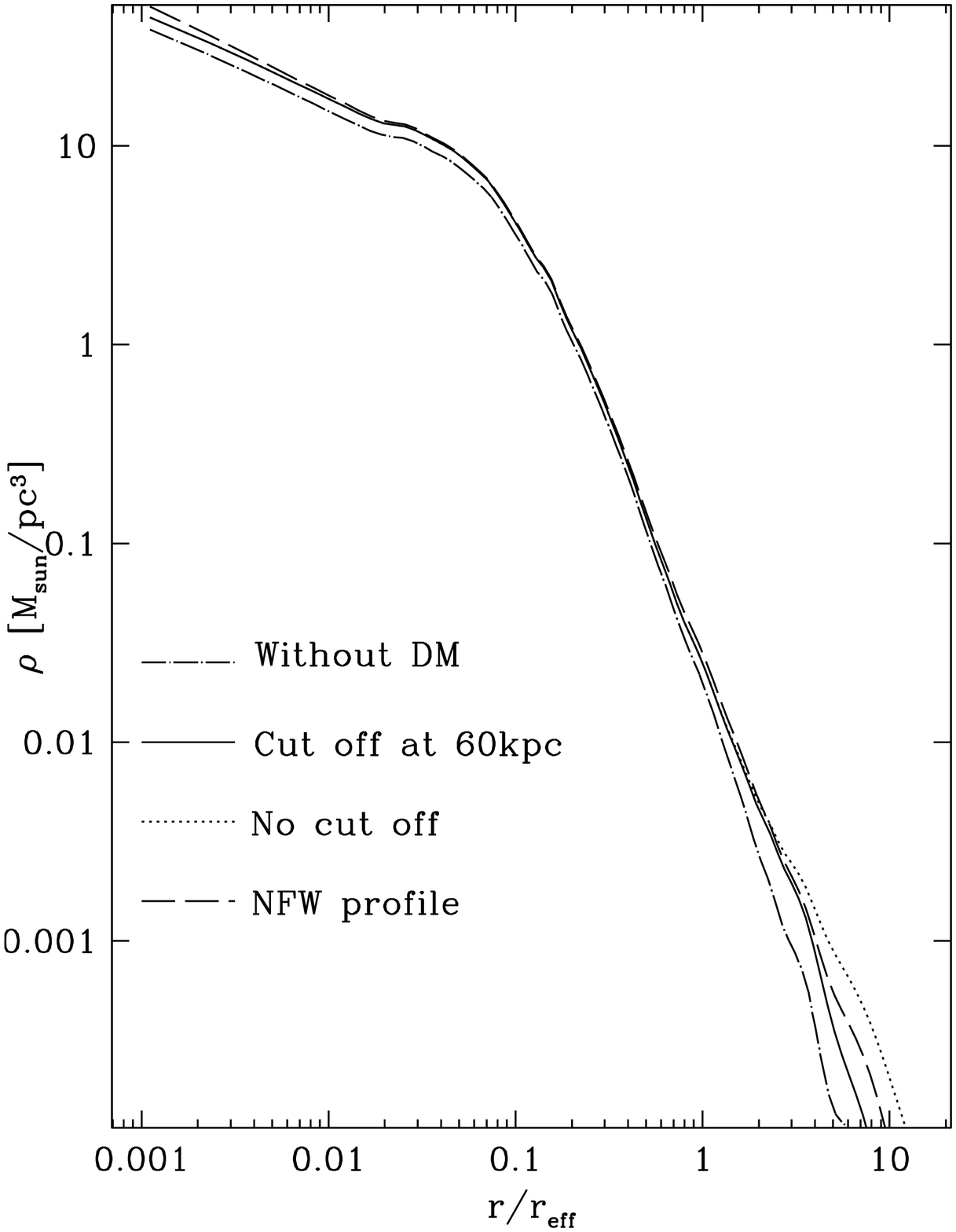}\\
\end{tabular}
\end{center}
\caption{The total density profile derived for \object{NGC1600} for the self-consistent 
model (dot-dashed line), the LOG dark matter halo cut at 60 kpc (full line), 
the LOG dark matter halo without cutoff (dotted line), and
the NFW dark matter halo (dashed line). 
  \label{fig_density}}
\end{figure}

Figure \ref{fig_density} shows the total density profile derived for
\object{NGC1600} for the different cases discussed above (self-consistent,
with a logarithmic dark matter halo with or without cut-off, or with a
NFW halo). Clearly, the $V_{esc}-Mgb$ correlation allows us to
constrain the outer shape of the total matter density, rather than
delivering a precise determination of the radius where the dark matter
halos terminate. In the next section we discuss this degeneracy in more detail.

\subsection{The degeneracy between outer halo slope and 
cut-off radius}
\label{sec_deg}

A steeper outer halo slope
results in less mass in the outer regions and, thus, a lower escape
velocity. Likewise, a lower cut-off radius reduces the outer mass and
the escape velocity as well. Consequently, a steeper density fall-off
in the outer halo region could be partly compensated for by a larger
cut-off radius.  In the following we investigate this degeneracy
quantitatively with spherical test models.

The test models are set up as follows. We assume that the galaxy's
light distribution can be described by a Hernquist sphere \citep{Her90}. For
simplicity we assume that the mass-to-light ratio equals one in
appropriate units. Then, the stellar mass distribution $\rho_\ast$
reads
\begin{equation}
\rho_\ast = \frac{M}{2 \pi} \frac{a}{r} \frac{1}{r(r+a)^3},
\end{equation}
where $M$ is the total stellar mass and $a$ is the scaling radius
\citep{Her90}.  In the following we use $M=3 \times 10^{11} \, M_\odot$
and $r_\mathrm{eff} = 10$ kpc ($r_\mathrm{eff} \approx 1.8 a$ for the
Hernquist sphere). To investigate the degeneracy between slope and
cut-off radius we seek a halo-profile whose slope can be adjusted
conveniently. For example
\begin{equation}
\label{rhodmtest}
\rho_\mathrm{DM} = \frac{\rho_0}{(r+r_0)^\gamma} \times \exp \left( 
-\frac{r^2}{r_\mathrm{cut}^2} \right).
\end{equation}
The four free parameters here are the normalisation $\rho_0$, a
scaling radius $r_0$ (inside which the halo density eventually becomes
constant), the cut-off radius $r_\mathrm{cut}$ and the asymptotic
logarithmic slope $\gamma$ (in the absence of a cut-off). For the rest
of this section we fix two of these parameters according to empirical
scaling relations for early-type galaxies. Firstly, the normalisation
$\rho_0$ is chosen such that
\begin{equation}
\rho_\ast(1.5 \, r_\mathrm{eff}) = \rho_\mathrm{DM}(1.5 \, r_\mathrm{eff})
\end{equation} 
\citep{Tho07}. Secondly, the halo scaling radius is fixed to 
$r_0 = 2.5 \, r_\mathrm{eff}$ \citep{Tho09}.

Since the goal is to investigate how much the outer slope is
degenerate with the cut-off radius we first construct an input
escape-velocity profile by fixing the remaining halo parameters
$\gamma$ and $r_\mathrm{cut}$ to some values, e.g. $\gamma=2$ and
$r_\mathrm{cut}=5\,r_\mathrm{eff} $.  We calculate the corresponding
escape velocity profile $V_{esc}$ from the total
potential\begin{equation} \Phi = \Phi_\ast + \Phi_\mathrm{DM},
\end{equation}
where $\Phi_\ast$ is given analytically for the Hernquist sphere
\citep{Her90} and $\Phi_\mathrm{DM}$ is obtained by numerically
integrating equation (\ref{eq_phispher}) for $\rho_\mathrm{DM}$ of
equation (\ref{rhodmtest}). We project the escape velocity via
Eq. \ref{eq_vescproj}. Note that the projection does not affect the
results discussed below. For the aim to study the degeneracy between
halo slope and cut-off radius we assume that the projected escape
velocity is given at 20 logarithmically spaced radii between $0.1 \,
r_\mathrm{eff}$ and $5 \, r_\mathrm{eff}$, with an uncertainty of
$\Delta \log V_{esc,pr} = 0.05$ (see Fig.\ref{fig_vesc}).

\begin{figure}
\begin{center}
\begin{tabular}{c}
\includegraphics[width=0.48\textwidth]{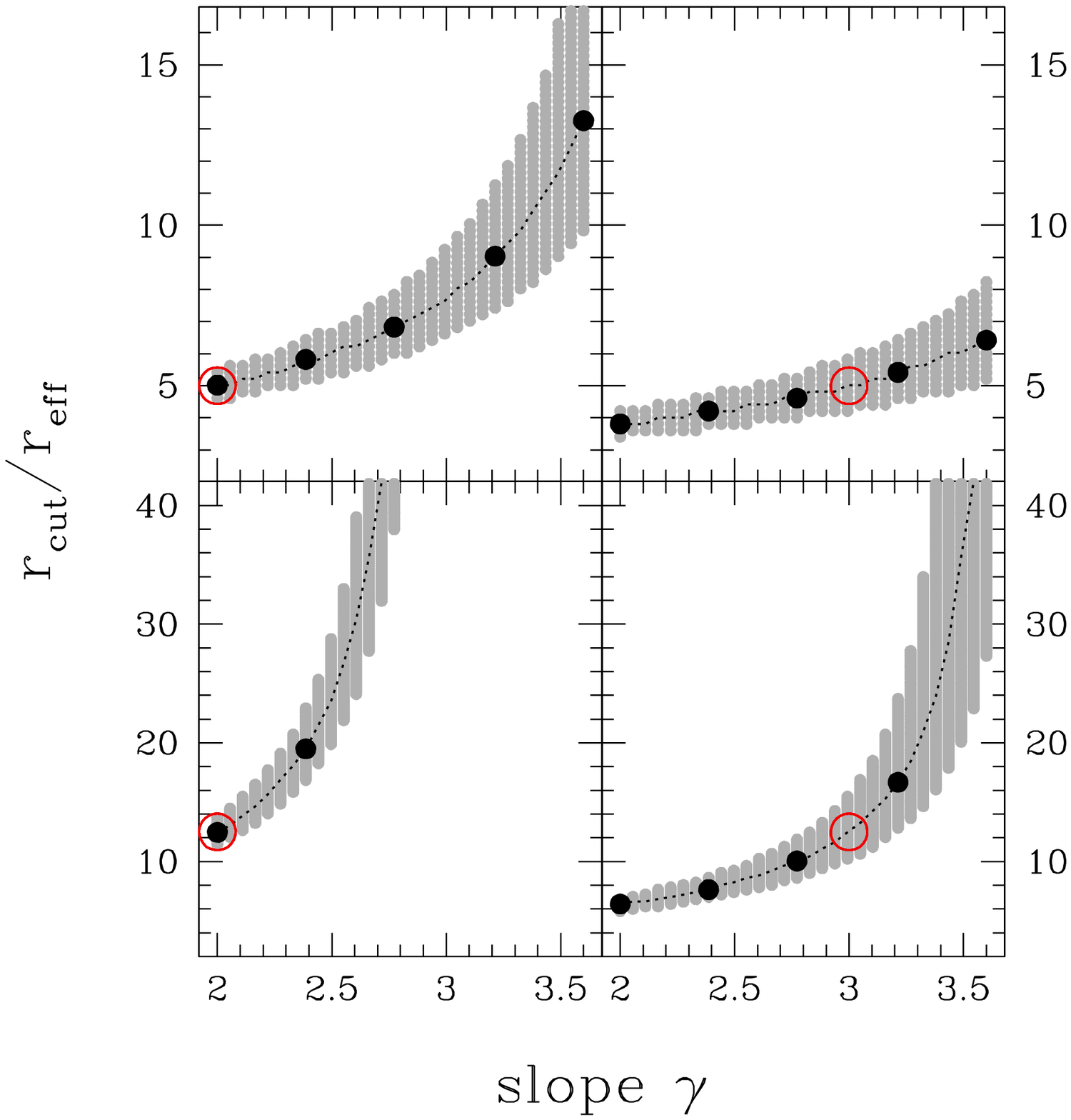}\\
\end{tabular}
\end{center}
\caption{Fitted cut-off radius $r_\mathrm{cut}$ against halo-slope. The input 
model is indicated by the red open circle. The filled black dots show the 
best-fit cut-off radius for halo-slopes $\gamma = 2,2.4,2.8,3.2,3.6$. Small 
grey dots indicate the region within $\Delta \chi^2 \le 1$ with respect to the 
input model. Left panels: input-slope $\gamma=2$ (isothermal-like); right 
panels: input-slope $\gamma = 3$ (NFW-like). Top row: input
$r_\mathrm{cut} = 5 \, r_\mathrm{eff}$; bottom row: input
$r_\mathrm{cut} = 12.5 \, r_\mathrm{eff}$.
  \label{r_c-slope}}
\end{figure}

Can we fit the so constructed escape-velocity profile also if we make
a wrong assumption upon the halo slope? To answer this question, the
slope is reset, e.g. $\gamma = 3$ and the remaining free parameter
$r_\mathrm{cut}$ is fitted such that the $\chi^2$ between the input
escape-velocity profile and the corresponding profile with the new
halo-slope is minimised. Fig.~\ref{r_c-slope} shows that for a wide
range of outer halo-slopes -- we probed $\gamma \in [2,3.6]$ -- the
escape-velocity profile can be reproduced well (with a $\chi^2 <
1$). As expected, a steeper halo-slope requires a larger cut-off
radius to compensate for the lowered outer mass-density. Our first
result is therefore that with the given uncertainties on the
escape-velocity there is a large uncertainty in the outer halo-slope
and cut-off radius, related to the degeneracy between the two.

\begin{figure}
\begin{center}
\begin{tabular}{c}
\includegraphics[width=0.48\textwidth]{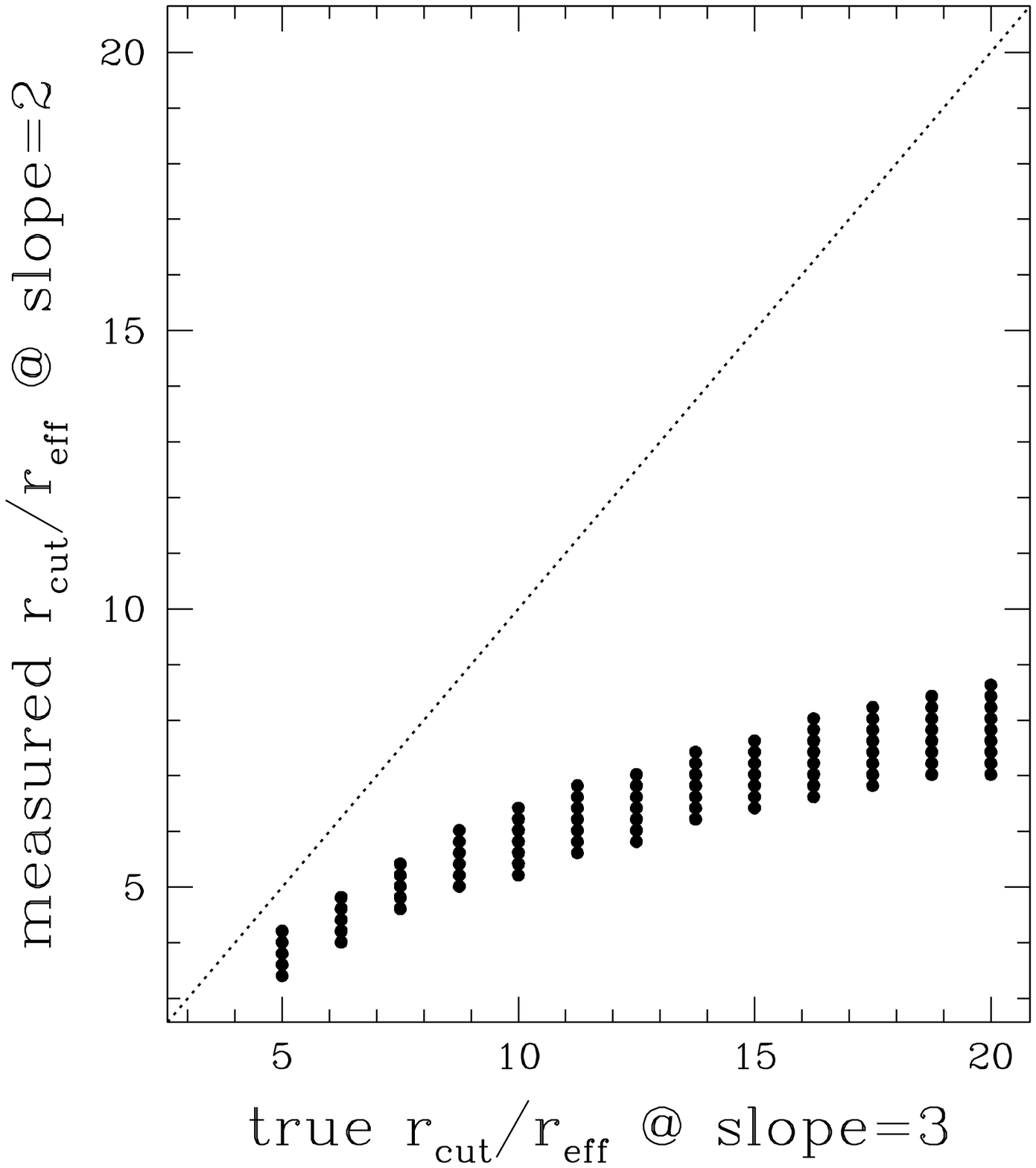}\\
\end{tabular}
\end{center}
\caption{Reconstructed halo cut-off radii (assuming an outer halo-slope 
$\gamma=2$) against true cut-off radii (the true halo-slope is $\gamma =3$). 
The points show the 68\% confidence region of the reconstructed cut-off radii. 
The dotted line is the one-to-one relation. An underestimation of the 
steepness of the halo density profile results in an underestimation of the 
cut-off radius.
\label{r_c-r_c}}
\end{figure}

In our standard modeling, the halo-slope is $\gamma \approx 2$
(cf. Eq. \ref{eq_log}).  Next we ask how much this assumption can
bias the 'measured' cut-off radii if the actual halo-slope is
steeper. To this extent we repeated the above analysis for a wide
range of input cut-off radii ($r_\mathrm{cut}/r_\mathrm{eff} \in
[5,20]$) and an input slope of $\gamma = 3$. The choice of $\gamma =
3$ is driven by the fact that pure dark matter cosmological $N$-body
simulations predict a halo profile close to Eq. (\ref{eq_NFW}),
i.e. $\gamma =3$. For each input model, we determined the 68\%
confidence region that would result under the assumption of $\gamma =
2$.  Fig.~\ref{r_c-r_c} shows that cut-off radii reconstructed with
$\gamma = 2$ are always too small -- as already discussed above. The
figure also shows that if the true cut-off radius is increased by a
factor of four, the reconstructed one (with $\gamma=2$) increases only
by factor of two. In other words, if the steepness of the halo density
profile is underestimated then (1) the cut-off radii are biased too
low and (2) a relatively small uncertainty in the 'measured' $r_\mathrm{cut}$ 
-- obtained under the assumption of a $\gamma = 2$ -- may 
correspond to a relatively large uncertainty in the true cut-off radii if
the real value of $\gamma$ is significantly larger than two.

\section{Discussion and summary}
\label{sec_summary}

Accurate kinematic profiles extending out to 1.5 to 2 $R_e$ along the
major and minor axis for the three giant elliptical galaxies \object{NGC1600},
\object{NGC4125} and NGC7169 have been measured, together with 6 lines
strength Lick indices. For \object{NGC4125} we also detected gas emission
along the major axis and measured its kinematics and equivalent width
strength. From the comparison of the
[NI]$\lambda\lambda$5197,5200/H$\beta$ vs [OIII]$\lambda$5007/H$\beta$
diagnostic diagram to the previous work, we concluded that the
emission region is probably caused by LINER-like emission.

With the help of the SSP models, we derived the stellar population
parameters, the M/L ratios and the broad band colors of our
galaxies. We found the galaxies \object{NGC1600} and \object{NGC7619} have high
metallicities, in contrast, the galaxy \object{NGC4125} is just slightly above
solar metallicity. The three objects have significant metallicity
gradients.  All of the galaxies are overabundant, \object{NGC4125} just by
$0.1$ dex, and do not have significant gradients of element abundances
along the axes.  These galaxies have sharp red peaks at the center,
reflecting the steep change of metallicity and in agreement with the
models predicted color profiles.

According to the simple element enrichment scenario, the $\alpha$
elements are mainly delivered by Type II supernovae explosions of
massive progenitor stars, and a substantial fraction of Fe peak
elements come from the delayed exploding Type Ia supernovae
\citep{1984ApJ...286..644N,1996ApJ...460..408T}. Thus the $\alpha/Fe$
can be used as an indicator to constrain the formation timescale of
stars. Hence the absence of radial variations of $\alpha$/Fe ratio in
our galaxies likely suggests that there is no radial variation of star
formation time scales.  The radial metallicity and lines strength
gradients give one of the most stringent constraints on the galaxy
formation. The galaxies that form monolithically have steeper
gradients and the galaxies that undergo major mergers have shallower
gradients.  The mean metallicity gradients for non-merger and merger
galaxies derived by theoretical simulation in
\citet{2004MNRAS.347..740K} are $\triangle$[Z/H]/$\triangle$ log(r)
$\sim -0.30 \pm 0.2$ and $-0.22 \pm 0.2$, respectively. The author
found that the galaxies with gradients steeper than -0.35 are all
non-major merger galaxies.  As it can be seen from Table
\ref{tab_grad} the gradients $\triangle [Z/H]/\triangle \log(r)$ of
our galaxies are compatible with numerical simulations.  At face
value, there is a weak indication that \object{NGC1600} and \object{NGC7619} were formed
through a monolithic collapse process, while \object{NGC4125} was shaped via a
(recent) major merger as also indicated by minor-axis rotation, the
presence of gas, young central stellar ages, just mild
$\alpha$/Fe-overabundance and the unrelaxed appearance of the outer
stellar envelope.

Using the axisymmetric Schwarzschild's orbits superposition technique
\citep{1979ApJ...232..236S} with and without dark halos, we derived
the local escape velocity of these galaxies.  From the correlation
between Mg\emph{b}, and the local escape velocity, we further
confirmed the suggestion of \citet{1990ApJ...359L..41F} that
metallicity and lines strength are a function of local escape
velocity. Moreover, we considered models with logarithmic dark matter
halos of different sizes. The best fitting Mg\emph{b}-$V_{esc}$
relation comes from models which cut off the logarithmic halo at 60
kpc. Similar cut-off radii are obtained when using the correlation
between $[Z/H]$ and $V_{esc}$, but the scatter in this relation is
larger, because it also depends on the stellar population age (see
also \citealt{2009MNRAS.398.1835S}).  We find that -- at a given
escape velocity -- the youngest of our galaxies, \object{NGC4125}, is also the
most metal-rich one, which could be explained within a merger scenario
as star formation from enriched material.

Our cut-off radii agree with the lensing analysis of cluster galaxies
by \citet{2007ApJ...656..739H} but not with the results from X-rays
and strong-lensing \citep{arXiv:0911.0678,2009ApJ...703L..51K}.
Larger cut-off radii are compatible with the data if dark matter halos
with steeper outer density profiles are considered.

We discover deviations from a linear correlation between log
Mg\emph{b} and log $V_{esc}$ in the outer parts of our galaxies, where
the measured Mg\emph{b} are lower than predicted by the extrapolation
from the correlation further inside. The physical link between
$V_{esc}$ and Mg\emph{b} is established at the main time of star
formation activity of the galaxy. Dry mergers happening after this
episode will increase $V_{esc}$ without increasing Mg\emph{b}.
Hydrodynamical cosmological simulations by \citet{2009ApJ...699L.178N}
indicate that the outer parts of massive ellipticals (beyond $\ga 5 \,
\mathrm{kpc}$) might be dominated by stars which were born in low-mass
halos and accreted by frequent minor mergers. If this would be the
dominant process for the formation of the outer parts of giant
ellipticals one would expect such a break-down of the linear
correlation between log Mg\emph{b} and log $V_{esc}$ as observed,
because the outer Mg\emph{b} would come from stars which were formed
in small halos with low $V_{esc}$ (and accordingly low Mg\emph{b}).

We plan to further investigate the connection
between line strengths and dark matter halos with a more extended galaxy sample.

\begin{acknowledgements}
We specially thank the McDonald Observatory for performing the
observations with Hobby-Eberly Telescope (HET) in service mode.  The
HET is a joint project of the University of Texas at Austin, the
Pennsylvania State University, Stanford University,
Ludwig-Maximilians-Universit\"{a}t M\"{u}nchen, and
Georg-August-Universit\"{a}t G\"{o}ttingen. The HET is named in
honor of its principal benefactors, William P. Hobby and Robert E.
Eberly." The Marcario Low Resolution Spectrograph is named for Mike
Marcario of High Lonesome Optics who fabricated several optics for
the instrument but died before its completion. The LRS is a joint
project of the Hobby-Eberly Telescope partnership and the Instituto
de Astronom\'{\i}a de la Universidad Nacional Aut\'{o}noma de
M\'{e}xico. This work was in part supported by the Chinese National
Science Foundation (Grant No. 10821061) and the National Basic
Research Program of China (Grant No. 2007CB815406). We also
gratefully acknowledge the Chinese Academy of Sciences and
Max-Planck-Institut f\"{u}r extraterrestrische Physik that
partially supported this work. Z.Han thanks the support of the
Chinese Academy of Sciences (Grant No. KJCX2-YW-T24).
Finally, we thank the referee, Paolo Serra, for a constructive report.
\end{acknowledgements}

\bibliographystyle{aa}
\bibliography{14395ref}


\end{document}